# Toward a Quantum Computing Formulation of the Electron Nuclear Dynamics Method via Fukutome Unitary Representation

Juan C. Domínguez[1], Ismael de Farias[2,++], and Jorge A. Morales[1,**]

[1]Department of Chemistry and Biochemistry, Texas Tech University,

Box 41061, Lubbock, TX 79409-1061, USA.

[2]Department of Industrial, Manufacturing and Systems Engineering, Texas Tech University,

Lubbock, TX 79430, USA.

**Corresponding author; *e - mail address*: jorge.morales@ttu.edu
++Deceased

**Abstract**:

We present the first installment of the quantum computing (QC) formulation of the electron nuclear dynamics (END) method within the variational quantum simulator (VQS) scheme: END/QC/VQS. END is a time-dependent, variational, on-the-flight, and non-adiabatic method to simulate chemical reactions. END represents nuclei with frozen Gaussian wave packets and electrons with a single-determinantal state in the Thouless non-unitary representation. Within the hybrid quantum/classical VQS, END/QC/VQS evaluates the metric matrix **M** and gradient vector **V** of the symplectic END/QC equations on a quantum computer, and calculates basis function integrals and time evolution on a classical computer. To adapt END to QC, we substitute the Thouless non-unitary representation with Fukutome unitary representation. We formulate the first END/QC/VQS version for pure electronic dynamics in chemical models consisting of two-electron units. Therein, Fukutome unitary matrices factorize into products of triads of one-qubit rotational matrices, which leads to a QC encoding of one electron per qubit. We design QC circuits to evaluate **M** and **V** in one-electron diatomic molecules. In $\log_2$-$\log_2$ plots, errors and deviations of those evaluations decrease linearly with the number of shots and with slopes = -1/2. We illustrate an END/QC/VQS simulation with the pure electronic dynamics of $H_2^+$. We discuss the present results and future END/QC/QVS extensions.

1. Introduction:

Quantum chemical dynamics seeks to describe chemical reactions in terms of quantum mechanics. Such an endeavor requires the solution of the time-dependent Schrödinger equation[1]: a partial linear differential equation, second-order in the particles' positions $\mathbf{r}_i$ and first order in time $t$, whose unknown is the time-dependent wavefunction $|\Psi(t)\rangle = \Psi(\mathbf{r}_i, s_i, t)$, where $s_i$ are the spin variables. Once $|\Psi(t)\rangle$ is found, all the chemical properties of a reactive system can be calculated from it. Unfortunately, the solution of the time-dependent Schrödinger equation becomes computationally expensive even for relatively small molecular systems. Therefore, some approximations should be introduced for feasibility's sake. One fruitful approach is to adopt the time-dependent variational principle (TDVP)[2]. Therein, a trial wavefunction $|\Psi(t)\rangle = |\Psi[\xi_i(t)]\rangle = \Psi[\mathbf{r}_i, s_i; \xi_i(t)]$ depending on a set of time-dependent parameters $\{\xi_i(t)\}$ is optimized for dynamical evolution by subjecting it to the stationary condition of the quantum action: $\delta A[\Psi^*[\xi_i(t)], \Psi[\xi_i(t)]] = 0, \forall\, \xi_i(t)$[2]. This procedure generates a system of classical-like equations of motion for the parameters $\{\xi_i(t)\}$ that takes the place of and is easier to integrate than the time-dependent Schrödinger equation. Furthermore, guided by chemical insight, one can propose relatively simple trial wavefunctions $|\Psi[\xi_i(t)]\rangle$ that deliver accurate predictions at low computational cost. A successful employment of the TDVP for quantum chemical dynamics has been provided by the electron nuclear dynamics (END) method[3,4]. END adopts a total trial wavefunction that contains frozen Gaussian wave packets to describe nuclei and a single-determinantal wavefunction in the Thouless representation[5] to describe electrons. END has proven to be both accurate and feasible for the simulation of a vast array of time-dependent chemical processes, e.g., Diels-Alder, $S_N2$, and ion cancer therapy reactions, *inter alia*[3,4,6-9].

While END/TDVP simulations run relatively fast, they become inevitably slower when applied to extremely large chemical systems, e.g., to the sizable biomolecules involved in ion cancer therapy reactions[4,6-9]. Therefore, we have devoted considerable efforts to accelerate the END/TDVP method for those simulations[4,6-9]. Broadly speaking, an END/TDVP simulation involves three essential computational tasks[3,4]: **(I)** The calculation of atomic and molecular basis functions integrals corresponding to the trial wavefunction, **(II)** the calculation of all the components of the END/TDVP equations of motion, and **(III)** the time integration of those equations. While tasks equivalent to I and III are present in other quantum chemistry methods, task II is specific to the END/TDVP approach and will be the main focus of this investigation. To accelerate the aforesaid tasks in our END/TDVP code PACE[4], we have employed various state-of-the-art techniques for *classical* (i.e. *standard*) computers, e.g., a mixed programming language (Python for logic flow and Fortran and $C^{++}$ for numerical calculations), intra- and internode parallelization, and the fast OED/ERD atomic integrals package[10] from the ACES[11] program, *inter alia.* Equipped with these capabilities, we could perform the first END/TDVP simulations of ion cancer therapy reactions involving large molecules, e.g., ion-induced water radiolysis in water clusters, and ion-induced DNA damage in nucleobases and the cytosine nucleotide[4,6-9]. Nevertheless, additional acceleration of END/TDVP simulations will be necessary to treat even larger systems.

In recent years, the burgeoning field of quantum computing (QC)[12-14] has shown an immense potential to revolutionize quantum chemistry through the provision of efficient *quantum computers* and *quantum algorithms* to simulate chemical systems. Thus, various research groups have reformulated established quantum chemistry methods[1] into the QC framework (cf. Refs.[15,16] and citations therein). In this context, Li and Benjamin have recently developed a variational quantum simulator (VQS) to calculate TDVP dynamics[17]. VQS is a *hybrid quantum/classical* approach wherein each computational task is entrusted to the type of computer, either *classical* or *quantum*, that provides the fastest algorithms with current technology. Specifically, the calculation of the trial wavefunction and its related components for the TDVP equations of motion are assigned to a quantum computer at each time step; those tasks are executed with quantum circuits that are modifications of one devised by Ekert *et al.*[18]. On the other hand, the time integration of the TDVP equations is assigned to a classical computer operating standard software. Li and Benjamin successfully applied their VQS to simulate the TDVP dynamics of an Ising-like model system, thus demonstrating the viability and potential of their innovative approach[17].

The VQS provides a paradigm for a hybrid quantum/classical implementation of TDVP dynamics, but it will remain scarcely relevant for quantum chemical dynamics if it is only applied to Ising-like models. While those models can simulate spin dynamics, they fail to reproduce most of the basic features of molecules and are, therefore, unsuitable for realistic simulations of chemical reactions. Therefore, the extension of the VQS approach for the accurate description of chemical reactions looms as a crucial endeavor in both quantum chemistry and QC fields. In this manuscript, we will embark into such an enterprise by reformulating the END method for QC (END/QC) within the VQS scheme (END/QC/VQS; henceforth, we will name END/QC the general QC formulation of END and END/QC/VQS its VQS realization).

In terms of the previously discussed tasks I-III, END/QC/VQS involves the following sequence of processes (cf. Fig. 1 for a flowchart). At a given time step, task I, the calculation of atomic and molecular basis functions integrals, is executed for the current nuclear and electronic configurations on a classical computer with the OED/ERD atomic integrals package[10]. With those integrals at hand, task II, the calculation of the components of the END/QC equations of motion, is executed on a quantum computer with quantum circuits that are END/QC versions of the general VQS circuit[17]. Specifically, our quantum circuits evaluate the END/QC trial wavefunction $|\Psi[\xi_i(t)]\rangle$, and the metric matrix $\mathbf{M}$ and energy gradient vector $\mathbf{V}$ of the END/QC equations of motion (cf. Sects. 2, 5-6). Then, task III, the time integration of the latter equations, is executed from the current to the next time step on a classical computer with standard differential equations solvers. The last task provides the new nuclear and electronic configurations for the next time step so that a new cycle of tasks I through III ensues; this loop runs from the initial to the final time of a dynamical simulation.

While transparent in outline, END/QC/VQS poses various theoretical and computational challenges during its development. As previously mentioned, the electronic part of the END trial wavefunction is a single-determinantal wavefunction in the Thouless representation[5]. In this approach, a *non-unitary* operator $\hat{Z}[z_i(t)]$ in terms of time-dependent parameters $z_i(t) \in \mathbb{C}$ generates the evolving electronic trial wavefunction $|\Psi[z_i(t)]\rangle_e$ by acting on a single-determinantal reference state $|0\rangle$: $|\Psi[z_i(t)]\rangle_e = \hat{Z}[z_i(t)]|0\rangle$ [3-5]. It is well-known that QC is strictly formulated in terms of unitary operators and gates[14]; therefore, the original END formulation in terms of a non-unitary operator $\hat{Z}[z_i(t)]$ does not directly fit into the unitary QC framework. To circumvent this difficulty (and also for pure theoretical reasons), we decided to reformulate END in terms of the *Fukutome unitary* representation[19] of a single-determinantal state. In this case, a *unitary* operator $\hat{U}[\lambda_i(t)]$ in terms of time-dependent parameters $\lambda_i(t) \in \mathbb{C}$ substitutes the previous non-unitary operator $\hat{Z}[z_i(t)]$ and generates the evolving electronic trial wavefunction as $|\Psi[\lambda_i(t)]\rangle_e = \hat{U}[\lambda_i(t)]|0\rangle$. In this way, $|\Psi[\lambda_i(t)]\rangle_e$ all its related terms fit directly into the unitary operators and gates of the quantum circuits. The aforesaid substitution of operators may seem simple in outline, but, in fact, it substantially changes the whole structure of the END formalism as will be shown in Sects. 3 and 5. While we will adhere to a unitary representation in this manuscript, we should note that a non-unitary END/QC formalism with the Thouless representation[5] is indeed possible if the involved non-unitary operators are expressed as linear combinations of unitaries (LCU)[20]. We will present that alternative formulation in a sequel.

The formulation of END/QC/VQS for all type of chemical systems is a challenging enterprise to be accomplished in stages. Therefore, in this first attempt, we will adopt some approximations and tackle particular systems. First, we will formulate END/QC for pure electronic dynamics in the presence of fixed nuclei. In this scheme, the END/QC equations of motion will explicitly involve electronic TVDP variables —nuclear variables will act implicitly in those equations as time-independent parameters. Second, for pure electronic dynamics, we will formulate END/QC for model systems described with effective Hamiltonians and minimal basis sets; similar types of systems are usually employed to develop and test QC methodologies for quantum chemistry[15,16]. During these formulations, we will emphasize the *continuous symmetry* aspects of the END/QC formalism within the Fukutome unitary representation[19]; i.e., the END/QC connection with the unitary Lie group and its associated Lie algebra. Finally, we will apply these END/QC developments to the simulation of pure electronic dynamics in one-electron diatomic molecules, paying special attention to the $H_2^+$ molecule. In the latter case, we will also consider the effect of the *point group spatial symmetry* in the END/QC equations of motions and dynamics.

As delineated in the previous paragraph, this manuscript presents the *proof of concept* of END/QC/VQS: A firm stepping stone from which we can continue developing this method to its full maturity. Thus, in upcoming publications, we will generalize the current END/QC/VQS for full electronic and nuclear dynamics, for general molecules described with *ab initio* Hamiltonians and large basis sets, and for executions on state-of-the-art quantum computers. Our group has presented a first glimpse into END/QC/VQS in Ref. [21]; however, herein, we present a more elaborated version of END/QC/VQS, both in terms of its formalism development and computer applications.

This manuscript is organized as follows. In Sect. 2, we will review the TDVP in terms of real variational parameters[2] because this form of the TDVP provides the most appropriate parameterization for the current version of END/QC/VQS. In Sect. 3, we will discuss Fukutome unitary representation[19] of single-determinantal wavefunctions in the context of the Hartree-Fock (HF) and the END methods. In Sect. 4, we will define a family of model chemical systems for END/QC/VQS treatments in the spirit of the semi-empirical methods[22] in quantum chemistry. In Sect. 5, we will formulate the END/QC formalism to simulate pure electronic dynamics in those systems. In Sect. 6, we will apply END/QC/VQS to simulate pure electronic dynamics in one-electron diatomic molecules, paying special attention to the $H_2^+$ molecule. Finally, in Sect. 7, we will analyze the main results of this investigation and discuss future work.

## 2. Background: Time-Dependent Variational Principle (TDVP):

There are different versions of the time-dependent variational principle such as the Dirac-Frenkel variational principle[23,24], the McLachlan variational principle[25], and the simply named TDVP (for their definitions and equivalence conditions, cf. Ref.[26-28]). Since END is based on the TDVP[3,4], we will employ this variational principle to formulate END/QC. The TDVP starts with a trial wavefunction $|\Psi(t)\rangle = |\Psi[\xi(t)]\rangle$ that depends on $N$ time-dependent variational parameters $\{\xi_j(t)\}$, $j = 1, \ldots N$, arranged in a column vector $\xi(t)$; these parameters can be real or complex[2-4], but, in this investigation, we will take them as real without any loss of generality: $\xi_j(t) \in \mathbb{R}$ $\forall j$ (for the relationship between real and complex TDVP parameterizations, cf. Ref.[2]). The selection of a trial wavefunction for a chemical problem is a matter of chemical insight; thus, following quantum chemistry experience, single-determinantal[3,4,29], multi-configuration[30], and coupled-cluster wavefunctions[31-33] have been employed/proposed to simulate chemical reactions in the TDVP framework. For a chosen $|\Psi[\xi(t)]\rangle$, the TDVP involves the quantum Lagrangian $L[\xi(t)]$ and quantum action $A[\xi(t)]$ functionals:

$$L[\xi(t)] = \frac{\langle \Psi[\xi(t)] | \frac{i}{2}\left(\frac{\vec{\partial}}{\partial t} - \frac{\overleftarrow{\partial}}{\partial t}\right) - \hat{H} | \Psi[\xi(t)] \rangle}{\langle \Psi[\xi(t)] | \Psi[\xi(t)] \rangle}; \quad A[\xi(t)] = \int_{t_1}^{t_2} L[\xi(t)] \, dt; \quad (1)$$

where $\hat{H}$ is the system Hamiltonian, $t_1$ and $t_2$ are the initial and final times of the dynamics, and $\vec{\partial}/\partial t$ and $\overleftarrow{\partial}/\partial t$ are time derivative operators acting to the right and to the left, respectively; these operators produce a real Lagrangian with a symmetric distribution of the derivatives $\dot{\xi}_j = d\xi_j/dt$ over the bra $\langle\ |$ and ket $|\ \rangle$ states[2]. The denominator of $L[\xi(t)]$ in Eq. (1) enforces normalization if the trial wavefunction is unnormalized[2]. However, we will parameterize $|\Psi[\xi(t)]\rangle$ with a unitary operator $\hat{U}[\xi(t)]$ acting on a reference state $|0\rangle$: $|\Psi[\xi(t)]\rangle = \hat{U}[\xi(t)]|0\rangle$, so that $\langle \Psi[\xi(t)] | \Psi[\xi(t)] \rangle = \langle 0 | \hat{U}[\xi(t)]^\dagger \hat{U}[\xi(t)] | 0 \rangle = 1$ at all times. Therefore, we will omit the aforesaid denominator in $L[\xi(t)]$ and in all the subsequent TDVP equations (for the TDVP

equations with an unnormalized trial wavefunction, cf. Refs. [2-4]). Under this condition, the Lagrangian $L(\boldsymbol{\xi})$ is[2]

$$L(\boldsymbol{\xi}) = \sum_{j=1}^{N} P_j(\boldsymbol{\xi}) \dot{\xi}_j - E(\boldsymbol{\xi});$$

$$E(\boldsymbol{\xi}) = \langle \Psi(\boldsymbol{\xi}) | \hat{H} | \Psi(\boldsymbol{\xi}) \rangle;\ P_j(\boldsymbol{\xi}) = i \left\langle \Psi(\boldsymbol{\xi}) \bigg| \frac{\partial \Psi(\boldsymbol{\xi})}{\partial \xi_j} \right\rangle = -i \left\langle \frac{\partial \Psi(\boldsymbol{\xi})}{\partial \xi_j} \bigg| \Psi(\boldsymbol{\xi}) \right\rangle; \quad (2)$$

where $E(\boldsymbol{\xi}) \in \mathbb{R}$ is the expectation value of the total energy, and $P_j(\boldsymbol{\xi}) \in \mathbb{R}$ is the canonical variable conjugate to $\xi_j$. To obtain the TDVP equations of motion, one should impose the stationary condition into the quantum action functional, $\delta A[\boldsymbol{\xi}(t)] = 0$, with respect to all the variational parameters $\boldsymbol{\xi}(t)$ and subjected to the end-point conditions $\delta \boldsymbol{\xi}(t_1) = \delta \boldsymbol{\xi}(t_2) = 0$. That procedure leads to a set of Euler-Lagrange equations for the parameters $\boldsymbol{\xi}(t)$:

$$\frac{d}{dt}\left(\frac{\partial L}{\partial \dot{\xi}_j}\right) = \left(\frac{\partial L}{\partial \xi_j}\right) \quad j = 1, \ldots N; \quad (3)$$

that in terms of the expressions in Eq. (2) are[2]:

$$\sum_{q=1}^{N} M_{pq}(\boldsymbol{\xi})\, \dot{\xi}_q = V_p \qquad p = 1, \ldots N$$

$$M_{pq}(\boldsymbol{\xi}) = i \frac{\partial \langle \Psi |}{\partial \xi_p} \frac{\partial | \Psi \rangle}{\partial \xi_q} + \text{H. c.} \ ;\ V_p = \frac{\partial E(\boldsymbol{\xi})}{\partial \xi_p} = \frac{\partial \langle \Psi(\boldsymbol{\xi}) |}{\partial \xi_p} \hat{H} | \Psi(\boldsymbol{\xi}) \rangle + \text{H. c.} \quad (4)$$

Above, the metric matrix $\mathbf{M} = (M_{pq})$ is real and antisymmetric, and the energy gradient vector $\mathbf{V} = (V_p)$ is real; $\mathbf{M}$ and $\mathbf{V}$ contain the kinematic and dynamic features of the system, respectively. It is clear from Eqs. (1)-(4) that the TDVP and its equations of motion are the quantum analogues of the classical Hamilton principle and of the classical Hamilton equations in *symplectic* form[34], respectively. In that scheme, the parameters $\boldsymbol{\xi}$ and $\mathbf{P}(\boldsymbol{\xi})$ span a generalized quantum phase space[2-4,35]. Solving Eq. (4) for the time-dependent parameters $\xi(t)$ provides the evolution of the trial wavefunction $|\Psi[\xi(t)]\rangle$ in time.

### 3. Fukutome Unitary Representation of Single-Determinantal Wavefunctions

As discussed in Sect. 1, we will develop the END/QC formalism employing Fukutome unitary representation[19] of single-determinantal states. Therefore, to understand our formalism, we should review such a unitary approach in the context of the HF theory[36] (for its extension to the Kohn-Sham density functional theory, cf. Ref. [29]). As anticipated in Sect. 1, we will consider a system containing $N_e$ evolving electrons and $N_N$ fixed nuclei. The electronic description of that system involves a set of $K$ orthonormal HF spin-orbitals $\{\psi_\zeta(\mathbf{x}_i)\} = \{\tilde{\psi}_\zeta(\mathbf{r}_i)\sigma_\zeta(s_i)\}$, $\zeta$, $i = 1, 2,$ ... $K > N_e$: $\langle \psi_\zeta | \psi_\eta \rangle = \int \psi_\zeta^*(\mathbf{x}_1) \psi_\eta(\mathbf{x}_1) d\mathbf{x}_1 = \delta_{\zeta\eta}$, where $\tilde{\psi}_\zeta(\mathbf{r}_i)$ and $\sigma_\zeta(s_i)$ are spatial orbitals and spin eigenfunctions with position and spin variables $\mathbf{r}_i$ and $s_i$, respectively, and $\mathbf{x}_i = (\mathbf{r}_i, s_i)$. Associated with the $\{\psi_\zeta\}$, we have a set of second-quantization creation $a_\zeta^\dagger$ and annihilation $a_\zeta$ operators that satisfy the anti-commutation relationships[36]:

$$\{a_\zeta^\dagger, a_\eta\} = \delta_{\zeta\eta}; \quad \{a_\zeta^\dagger, a_\eta^\dagger\} = \{a_\zeta, a_\eta\} = 0. \tag{5}$$

Having the vacuum state $|vac\rangle$, $a_\zeta|vac\rangle = 0 \; \forall \zeta$, we can define a reference Slater determinant state $|0\rangle$ with occupied spin-orbitals $\psi_\alpha, \psi_\beta \ldots \psi_{N_e}$ as:

$$|0\rangle = a_\alpha^\dagger a_\beta^\dagger \ldots a_N^\dagger |vac\rangle = |\psi_\alpha \psi_\beta \ldots \psi_{N_e}\rangle = \det[\ldots \psi_\alpha(\mathbf{x}_i)\ldots]. \tag{6}$$

In this investigation, we will follow Fukutome's notation[19] and denote the $N_e$ occupied (hole) spin-orbitals in $|0\rangle$ with the indices $\alpha, \beta, \gamma...$, the $K - N_e$ unoccupied (particle) spin-orbitals

with the indices $\mu$, $\nu$, $\xi$ ..., and the whole $K$ spin-orbitals irrespective of their with occupancy with the indices $\varsigma$, $\eta$, $\iota$ ... From $|0\rangle$, we can generate all the remaining single-, double-, etc., -excitation Slater determinants as $|\Psi_\alpha^\mu\rangle = a_\mu^\dagger a_\alpha |0\rangle = |\psi_\mu \psi_\beta ... \psi_N\rangle$, $|\Psi_{\alpha\beta}^{\mu\nu}\rangle = a_\nu^\dagger a_\beta a_\mu^\dagger a_\alpha |0\rangle = |\psi_\mu \psi_\nu ... \psi_N\rangle$, etc. All these single-determinantal states are orthonormal among themselves. While $|0\rangle$ is arbitrary, we will take it as the HF ground state as is usually the case in END simulations[3,4]. The electronic Hamiltonian $\hat{H}_e$ of the system in second-quantization form is

$$\hat{H}_e = \sum_{\varsigma,\eta} h_{\varsigma\eta} a_\varsigma^\dagger a_\eta + \frac{1}{2} \sum_{\varsigma,\eta,\kappa,\iota} \langle \varsigma\eta | \kappa\iota \rangle a_\varsigma^\dagger a_\eta^\dagger a_\iota a_\kappa; \tag{7}$$

where $h_{\varsigma\eta}$ and $\langle \varsigma\eta | \kappa\iota \rangle$ are the one- and two-electron integrals from the spin-orbitals $\{\psi_\varsigma\}$ [36].

To formulate the HF theory in unitary form, Fukutome considered the set of $K^2$ pair operators $a_\eta^\dagger a_\varsigma$ that span (generate) the $U(K)$ Lie algebra (group) with commutation relationships[37-39]:

$$[a_\varsigma^\dagger a_\eta, a_\iota^\dagger a_\kappa] = \delta_{\eta\iota} a_\varsigma^\dagger a_\kappa - \delta_{\varsigma\kappa} a_\iota^\dagger a_\eta; \tag{8}$$

From them, one can construct operator $\hat{U}[\mathbf{u}(\boldsymbol{\gamma})]$ and matrix $\mathbf{u}(\boldsymbol{\gamma})$ realizations of the unitary Lie group $U(K)$[37-39] as:

$$\hat{U}[\mathbf{u}(\boldsymbol{\gamma})] = \exp[\hat{\Gamma}(\boldsymbol{\gamma})] = \exp\left(\sum_{\varsigma,\eta} \gamma_{\varsigma\eta} a_\varsigma^\dagger a_\eta\right); \quad \mathbf{u}(\boldsymbol{\gamma}) = \exp(\boldsymbol{\gamma}); \quad \boldsymbol{\gamma} = (\gamma_{\varsigma\eta}) \in \mathbb{C}^{K\times K}; \quad \gamma_{\varsigma\eta} = -\gamma_{\eta\varsigma}^*; \tag{9}$$

where $\boldsymbol{\gamma} = (\gamma_{\varsigma\eta}) \in \mathbb{C}^{K\times K}$ is an anti-Hermitian matrix containing the parameters $\gamma_{\varsigma\eta} \in \mathbb{C}$. The unitary matrix $\mathbf{u}(\boldsymbol{\gamma}) = (u_{\varsigma\eta})$ acts on the orthonormal spin-orbitals $\{\psi_\varsigma\}$ as:

$$\left(\phi_1 \ldots \phi_\zeta \ldots \phi_K\right) = \left(\psi_1 \ldots \psi_\zeta \ldots \psi_K\right)\mathbf{u}(\boldsymbol{\gamma}) \Rightarrow \phi_\zeta = \sum_\eta \psi_\eta u_{\eta\zeta}; \tag{10}$$

where $\{\phi_\zeta(\mathbf{x}_i)\}$ is a new set of $K$ transformed spin-orbitals; since $\mathbf{u}(\boldsymbol{\gamma})$ is unitary, the $\{\phi_\zeta(\mathbf{x}_i)\}$ are orthonormal as well. Then, all the possible single-determinantal states $|\Psi(\boldsymbol{\gamma})\rangle$ that are non-orthogonal to $|0\rangle$, $\langle\Psi(\boldsymbol{\gamma})|0\rangle \neq 0$, can be generated from $|0\rangle$ by the unitary transformation[19]:

$$|\Psi(\boldsymbol{\gamma})\rangle = \hat{U}[\mathbf{u}(\boldsymbol{\gamma})]|0\rangle = |\phi_\alpha \phi_\beta \ldots \phi_{N_e}\rangle = \det[\ldots \phi_\alpha(\mathbf{x}_i)\ldots]; \tag{11}$$

where $|\Psi(\boldsymbol{\gamma})\rangle$ contains $N_e$ occupied (hole) transformed spin-orbitals $\{\phi_\alpha(\mathbf{x}_i)\}$.

Eq. (11) provides a unitary representation of the single-determinantal states $|\Psi(\boldsymbol{\gamma})\rangle$ but its application is inconvenient due to redundancies in the parameters $\boldsymbol{\gamma}$ as shown shortly. Fortunately, Fukutome[19] solved this problem via an exact factorization of both $\hat{U}[\mathbf{u}(\boldsymbol{\gamma})]$ and $\mathbf{u}(\boldsymbol{\gamma})$ into two unitary components containing the non-redundant and redundant parameters $\boldsymbol{\lambda}$ and $\boldsymbol{\xi}$, respectively:

$$\hat{U}[\mathbf{u}(\boldsymbol{\gamma})] = \hat{U}[\mathbf{u}(\boldsymbol{\lambda})]\hat{U}[\mathbf{u}(\boldsymbol{\xi})] \equiv \exp(\hat{\Lambda})\exp(\hat{\Xi}); \quad \mathbf{u}(\boldsymbol{\gamma}) = \mathbf{u}(\boldsymbol{\lambda})\mathbf{u}(\boldsymbol{\xi}) \equiv \exp(\boldsymbol{\lambda})\exp(\boldsymbol{\xi}); \tag{12}$$

where the new operators and complex parameters are

$$\hat{\Xi} = \sum_{\alpha,\beta} \xi_{\alpha\beta} a_\alpha^\dagger a_\beta + \sum_{\mu,\nu} \bar{\xi}_{\mu\nu} a_\mu^\dagger a_\nu; \quad \xi_{\alpha\beta} = -\xi_{\beta\alpha}^*; \quad \bar{\xi}_{\mu\nu} = -\bar{\xi}_{\nu\mu}^*; \quad \hat{\Lambda} = \sum_{\mu,\alpha}\left(\lambda_{\mu\alpha} a_\mu^\dagger a_\alpha - \lambda_{\mu\alpha}^* a_\alpha^\dagger a_\mu\right). \tag{13}$$

Notice that $\hat{\Xi}$ only contains hole-hole and particle-particle pair operators, and $\hat{\Lambda}$ only contains particle-hole and hole-particle ones. The matrix $\mathbf{u}(\boldsymbol{\xi})$ is $\mathbf{u}(\boldsymbol{\xi}) = \mathbf{w} \oplus \bar{\mathbf{w}}$ where $\mathbf{w} = \exp[(\xi_{\alpha\beta})] \in U(N_e)$ and $\bar{\mathbf{w}} = \exp[(\bar{\xi}_{\mu\nu})] \in U(K - N_e)$. The matrix $\mathbf{u}(\boldsymbol{\lambda}) \in U(K)$ is

$$\mathbf{u}(\boldsymbol{\lambda}) = \begin{bmatrix} \mathbf{C}(\boldsymbol{\lambda}) & -\mathbf{S}^\dagger(\boldsymbol{\lambda}) \\ \mathbf{S}(\boldsymbol{\lambda}) & \tilde{\mathbf{C}}(\boldsymbol{\lambda}) \end{bmatrix}; \quad \boldsymbol{\lambda} = (\lambda_{\mu\alpha}) \in \mathbb{C}^{(K-N) \times N}; \tag{14}$$

where

$$\mathbf{S}(\boldsymbol{\lambda}) = \sum_{k=0}^{\infty} \frac{(-1)^k}{(2k+1)!} \boldsymbol{\lambda}(\boldsymbol{\lambda}^\dagger \boldsymbol{\lambda})^k \in \mathbb{C}^{(K-N) \times N}; \quad \mathbf{C}(\boldsymbol{\lambda}) = \mathbf{I}_{N \times N} + \sum_{k=1}^{\infty} \frac{(-1)^k}{(2k)!} (\boldsymbol{\lambda}^\dagger \boldsymbol{\lambda})^k \in \mathbb{C}^{N \times N};$$

$$\tilde{\mathbf{C}}(\boldsymbol{\lambda}) = \mathbf{I}_{(K-N) \times (K-N)} + \sum_{k=1}^{\infty} \frac{(-1)^k}{(2k)!} \boldsymbol{\lambda}(\boldsymbol{\lambda}\boldsymbol{\lambda}^\dagger)^k \in \mathbb{C}^{(K-N) \times (K-N)}. \tag{15}$$

The functions $\mathbf{S}(\boldsymbol{\lambda})$ and $\mathbf{C}(\boldsymbol{\lambda})/\tilde{\mathbf{C}}(\boldsymbol{\lambda})$ are matrix generalizations of the standard sine and cosine functions, respectively. From a QC standpoint, $\mathbf{u}(\boldsymbol{\lambda}) \in U(K)$ in Eq. (14) can be interpreted as a multi-qubit generalization of the one-qubit matrices $\mathbf{u}(\phi_0, \phi_1, \phi_2, \phi_3) \in U(2)$ in terms of the rotational matrices $\mathbf{R}_i(\phi_i)$, $i = 1, 2, 3$, [14]; suggestively, $\mathbf{u}(\boldsymbol{\lambda})$ resembles $\mathbf{R}_y(\phi_y)$, cf. Sect. 5. The unitary matrices $\mathbf{w}$ and $\overline{\mathbf{w}}$ act on the spin-orbitals $\{\psi_\zeta\}$ as $\phi_\alpha = \sum_\beta \psi_\beta w_{\beta\alpha}$ and $\phi_\mu = \sum_\nu \psi_\nu \overline{w}_{\nu\mu}$, i.e., $\mathbf{w}$ combines occupied spin-orbitals $\{\psi_\beta\}$ among themselves, and $\overline{\mathbf{w}}$ combines unoccupied spin-orbitals $\{\psi_\nu\}$ among themselves. The matrix $\mathbf{u}(\boldsymbol{\lambda})$ acts on the spin-orbitals $\{\psi_\zeta\}$ as:

$$\phi_\alpha = \sum_\beta \psi_\beta [\mathbf{C}(\boldsymbol{\lambda})]_{\beta\alpha} + \sum_\mu \psi_\mu [\mathbf{S}(\boldsymbol{\lambda})]_{\mu\alpha}; \quad \phi_\mu = \sum_\nu \psi_\nu [\tilde{\mathbf{C}}(\boldsymbol{\lambda})]_{\nu\mu} - \sum_\alpha \psi_\alpha [\mathbf{S}^\dagger(\boldsymbol{\lambda})]_{\alpha\mu}; \tag{16}$$

i.e., $\mathbf{u}(\boldsymbol{\lambda})$ combines occupied and unoccupied spin-orbitals $\{\psi_\beta, \psi_\mu\}$ among themselves. From a QC standpoint, the $K$ transformed orthonormal spin-orbitals $\{\phi_\alpha, \phi_\mu\}$ in Eq. (16) can be interpreted as multiple-qubit generalization of the one-qubit Bloch sphere states[14], cf. Sect. 5. The action of the operator $\hat{U}[\mathbf{u}(\boldsymbol{\xi})]$ on $|0\rangle$ is

$$|\Psi(\xi)\rangle = \hat{U}\left[\mathbf{u}(\xi)\right]|0\rangle = \left|...\left(\sum_{\beta}\psi_{\beta}w_{\beta\alpha}\right)...\right\rangle = \det(\mathbf{w})|0\rangle \sim |0\rangle; \tag{17}$$

i.e., $\hat{U}\left[\mathbf{u}(\xi)\right]$ combines the occupied spin-orbitals $\{\psi_{\beta}\}$ in $|0\rangle$ among themselves and transforms $|0\rangle$ into the *equivalent* states $\det(\mathbf{w})|0\rangle$. The action of the operator $\hat{U}\left[\mathbf{u}(\lambda)\right]$ on $|0\rangle$ is

$$|\Psi(\lambda)\rangle = \hat{U}\left[\mathbf{u}(\lambda)\right]|0\rangle = \left|...\{\psi_{\beta}\left[\mathbf{C}(\lambda)\right]_{\beta\alpha} + \psi_{\mu}\left[\mathbf{S}(\lambda)\right]_{\mu\alpha}\}...\right\rangle = |...\phi_{\alpha}...\rangle; \tag{18}$$

i.e., $\hat{U}\left[\mathbf{u}(\lambda)\right]$ combines occupied and unoccupied spin-orbitals $\{\psi_{\beta}, \psi_{\mu}\}$ into $|0\rangle$ and transforms the latter into the *non-equivalent* states $|\Psi(\lambda)\rangle$. Thus, Eqs. (17) and (18) demonstrate that $\lambda$ and $\xi$ are the non-redundant and redundant parameters in the unitary representation of all the single-determinantal states from $|0\rangle$. Then, we can omit $\hat{U}\left[\mathbf{u}(\xi)\right]$ in $|\Psi(\gamma)\rangle = \hat{U}\left[\mathbf{u}(\gamma)\right]|0\rangle = \hat{U}\left[\mathbf{u}(\lambda)\right]\hat{U}\left[\mathbf{u}(\xi)\right]|0\rangle$, Eq. (11), and take $|\Psi(\lambda)\rangle = \hat{U}\left[\mathbf{u}(\lambda)\right]|0\rangle$, Eq. (18), as the non-redundant Fukutome unitary representation of the all the non-equivalent single-determinantal states $|\Psi(\lambda)\rangle$ from $|0\rangle$ and non-orthogonal to it[19].

Fukutome unitary representation, Eq. (18), establishes a bijective mapping between the single-determinantal states $|\Psi(\lambda)\rangle$ and the parameters $\lambda$ [3,4,19]. In a time-independent context, this representation provides a suitable parameterization of stationary states $|\Psi(\lambda)\rangle$ for HF energy optimizations, instabilities analyses, and symmetry breaking classifications[19]. In a time-dependent context, Fukutome unitary representation provides a suitable parameterization of time-evolving states $|\Psi[\lambda(t)]\rangle$ for END/QC optimizations and for time-dependent symmetry breakings, as the Thouless non-unitary representation[5] does in the standard END[3,4,40,41]. We will employ

Fukutome unitary representation to formulate END/QC in Sects. 5 and 6. In this and the following section, we keep Fukutome's original parameters $\gamma \in \mathbb{C}^{K \times K}$ but we will turn to real parameters in Sect. 5.

## 4. Model Systems for END/QC

The general formulation and code implementation of END/QC for any type of chemical system are challenging. We will present such a general treatment in a subsequent publication. Herein, as a *proof of concept*, we will implement END/QC for a family of model chemical systems, which in some cases admit analytical solutions. This family of models is constructed by a set of five approximations that resemble those employed in semi-empirical methods for quantum chemistry[22]. The END/QC trial wavefunction $|\Psi(\lambda)\rangle$ for any of these systems will be a single-determinantal wavefunction in Fukutome unitary representation, Eq. (18). The approximations defining these systems are:

**Approximation 1:** We adopt a system consisting of $\tilde{N}$ chemical units; each unit has two electrons with opposite spins up $\uparrow$ and down $\downarrow$, respectively. The units can be atoms, molecules, or monomers, identical or not. The units can be real subsystems with two actual electrons (e.g. H$^-$, Li$^+$, H$_2$, HeH$^+$, etc.) or model subsystems with two active electrons in the presence of a frozen electronic core[22]. The units together can constitute a single molecule (H$_2$), a super-molecule (H$_2$)$_n$, or a polymer/lattice (-- H$_2$ -- H$_2$ -- H$_2$ --).

**Approximation 2:** Within a unit, each electron has available two spin-orbitals, one occupied and another unoccupied with respect to the reference state. Specifically, the spin-orbitals $\{\psi_\alpha, \psi_\mu\}$ and $\{\psi_\beta, \psi_\nu\}$ are available for the unit electrons with spins up and down, respectively. This approximation implies a minimal basis set[36] with two atomic orbitals per unit to construct the spin-orbitals.

**Approximation 3:** Particle-hole excitations are only allowed within spin-orbitals with the same type of spin. To enforce this, we set in the END/QC trial wavefunction $|\Psi(\lambda)\rangle$, Eq. (18), and resulting expressions:

$$\lambda_{\mu\alpha} = 0 \qquad \text{if } \langle\alpha|\hat{s}_z|\alpha\rangle \neq \langle\mu|\hat{s}_z|\mu\rangle; \qquad (19)$$

where $\hat{s}_z$ is the z-component of the one-electron spin operator. With Eq. (19), $|\Psi(\lambda)\rangle$ becomes an unrestricted HF (UHF) state[36] [an axial-spin-density-wave (ASDW) wavefunction in Fukutome's classification of HF states[19]], where the transformed spin-orbitals $\phi_\alpha$ remain as one-electron spin eigenfunctions, and the spin symmetries of $|\Psi(\lambda)\rangle$ with respect to $\hat{S}_z$ and $\hat{S}^2$ (squared total spin) are preserved and lost, respectively. UHF wavefunctions are regularly used in quantum chemistry, especially to describe radicals and bond making/breaking processes[19,36]. Without this approximation, $|\Psi(\lambda)\rangle$ will be a less useful generalized HF (GHF) wavefunction[1,19], where the transformed spin-orbitals $\phi_\alpha$ contain both spin-up and spin-down components, and all the spin symmetries of $|\Psi(\lambda)\rangle$ are lost.

**Approximation 4:** Particle-hole excitations are only allowed within units; e.g., an electron in the occupied spin-up spin-orbital $\psi_\alpha$ can only be excited to the unoccupied spin-up spin-orbital $\psi_\mu$ in the same unit:

$$\lambda_{\mu\alpha} = 0 \qquad \text{if unit of } \psi_\alpha \neq \text{unit of } \psi_\mu. \qquad (20)$$

This approximation becomes exact with well-separated units carrying localized spin-orbitals.

**Approximation 5:** If Approximation 1 involves active electrons in the presence of a frozen electronic core, the *ab initio* Hamiltonian $\hat{H}_e$, Eq. (7), should be transformed into an effective Hamiltonian by applying appropriate semi-empirical approximations[22] to the one- and two-electron integrals of $\hat{H}_e$. Unlike the previous approximations, this one eludes a general formulation because it depends on the chemical features of the model system at hand. Therefore, we will present examples of this approximation as we investigate concrete systems in Sect. 6 of this manuscript and in subsequent publications.

## 5. Formulation of END/QC for the Model Systems. Natural QC Encoding

Application of approximations 1-4 in conjunction with the $U(K)$ commutation relationships, Eq. (8), to the operator $\hat{U}[\mathbf{u}(\boldsymbol{\lambda})]$ and matrix $\mathbf{u}(\boldsymbol{\lambda})$, Eqs. (12) and (14), leads to their factorization into $\mu-\alpha$ particle-hole-pair components:

$$\hat{U}[\mathbf{u}(\boldsymbol{\lambda})] = \prod_{(\mu\alpha)}^{N_e} \hat{U}[\mathbf{u}(\lambda_{\mu\alpha})]; \quad \mathbf{u}(\boldsymbol{\lambda}) = \otimes_{(\mu\alpha)}^{N_e} \mathbf{u}(\lambda_{\mu\alpha}); \tag{21}$$

where $\prod_{(\mu\alpha)}^{N_e}$ and $\otimes_{(\mu\alpha)}^{N_e}$ denote ordinary and tensor products, respectively, and $(\mu\alpha)$ is a compound index, $(\mu\alpha)=1, 2, \ldots N_e$. $\hat{U}[\mathbf{u}(\lambda_{\mu\alpha})]$ and $\mathbf{u}(\lambda_{\mu\alpha}) \in$ SU(2) depend on a single parameter $\lambda_{\mu\alpha} = \rho_{\mu\alpha} e^{+i\omega_{\mu\alpha}} \in \mathbb{C}$ as

$$\hat{U}[\mathbf{u}(\lambda_{\mu\alpha})] = \exp(\lambda_{\mu\alpha} a_\mu^\dagger a_\alpha - \lambda_{\mu\alpha}^* a_\alpha^\dagger a_\mu); \tag{22}$$

and

$$\mathbf{u}(\lambda_{\mu\alpha}) = \mathbf{R}(\rho_{\mu\alpha}, \omega_{\mu\alpha}) = \begin{bmatrix} C(\lambda_{\mu\alpha}) & -S^*(\lambda_{\mu\alpha}) \\ S(\lambda_{\mu\alpha}) & C(\lambda_{\mu\alpha}) \end{bmatrix} = \begin{bmatrix} \cos(\rho_{\mu\alpha}) & -e^{-i\omega_{\mu\alpha}} \sin(\rho_{\mu\alpha}) \\ e^{+i\omega_{\mu\alpha}} \sin(\rho_{\mu\alpha}) & \cos(\rho_{\mu\alpha}) \end{bmatrix}. \tag{23}$$

In Eqs. (21)-(23), the multi-qubit matrix $\mathbf{u}(\lambda) \in U(K)$, Eq. (14), simplifies into a tensor product of one-qubit unitary matrices $\mathbf{u}(\lambda_{\mu\alpha}) \in U(2)$ [14], the matrix $\lambda$, Eq. (14), into a set of numbers $\lambda_{\mu\alpha} = \rho_{\mu\alpha} e^{+i\omega_{\mu\alpha}}$, and the matrix functions $\mathbf{C}(\lambda)$, $\tilde{\mathbf{C}}(\lambda)$, and $\mathbf{S}(\lambda)$, Eq. (15), into the numerical functions $C(\lambda_{\mu\alpha})$ and $S(\lambda_{\mu\alpha})$ in terms of sine, cosine, and exponential functions. Furthermore, through $\lambda_{\mu\alpha} = \rho_{\mu\alpha} e^{+i\omega_{\mu\alpha}}$, we can switch from a complex $(\lambda_{\mu\alpha})$ to a real $(\rho_{\mu\alpha}, \omega_{\mu\alpha})$ parameterization, i.e., $\mathbf{u}(\lambda_{\mu\alpha}) = \mathbf{R}(\rho_{\mu\alpha}, \omega_{\mu\alpha})$ in Eq. (23); the latter scheme fits into the TDVP with real variational parameters discussed in Sect. 2.

The factorizations in Eq. (21) into particle-hole-pair components $(\mu\alpha)$ suggests a *natural QC encoding* for the current model systems. In this scheme, each electron of a $\mu - \alpha$ particle-hole-excitation pair can be assigned to a single qubit with compound index $(\mu\alpha)$. We will employ this QC encoding during the rest of this manuscript and present alternative formulations in terms of standard QC encodings (e.g. Jordan-Wigner, Kitaev, etc. [16]) in a sequel. In the natural QC encoding, the occupied and unoccupied spin-orbitals $|\psi_\alpha\rangle$ and $|\psi_\mu\rangle$ correspond to the computational basis states (vectors) $|0_\alpha\rangle$ and $|1_\mu\rangle$ of the qubit with index $(\mu\alpha)$:

$$\psi_\alpha \leftrightarrow |0_\alpha\rangle = \begin{pmatrix} 1 \\ 0 \end{pmatrix}_\alpha ; \quad \psi_\mu \leftrightarrow |1_\mu\rangle = \begin{pmatrix} 0 \\ 1 \end{pmatrix}_\mu ; \qquad (24)$$

where the equivalences between operator-wavefunction and matrix-vector representations are indicated with the symbol $\leftrightarrow$. In addition, we can define pseudo spin angular momentum operators $\hat{J}_{X\mu\alpha}$, $\hat{J}_{Y\mu\alpha}$, and $\hat{J}_{Z\mu\alpha}$, and an electron number operator $\hat{N}_{\mu\alpha}$ as

$$\hat{J}_{X\mu\alpha} = \frac{1}{2}\left(a_\mu^\dagger a_\alpha + a_\alpha^\dagger a_\mu\right) \leftrightarrow \frac{1}{2}\mathbf{X}_{\mu\alpha}; \quad \hat{J}_{Y\mu\alpha} = \frac{i}{2}\left(a_\mu^\dagger a_\alpha - a_\alpha^\dagger a_\mu\right) \leftrightarrow \frac{1}{2}\mathbf{Y}_{\mu\alpha};$$
$$\hat{J}_{Z\mu\alpha} = \frac{1}{2}\left(a_\alpha^\dagger a_\alpha - a_\mu^\dagger a_\mu\right) \leftrightarrow \frac{1}{2}\mathbf{Z}_{\mu\alpha}; \quad \hat{N}_{\mu\alpha} = a_\alpha^\dagger a_\alpha + a_\mu^\dagger a_\mu \leftrightarrow \mathbf{I}_{\mu\alpha};$$
(25)

where their equivalences with the identity and Pauli matrices $\mathbf{I}_{\mu\alpha}$, $\mathbf{X}_{\mu\alpha}$, $\mathbf{Y}_{\mu\alpha}$, and $\mathbf{Z}_{\mu\alpha}$ of the qubit $(\mu\alpha)$ are shown. By inverting the relationships in Eq. (25), one can encode the operators $a_\eta^\dagger a_\zeta$ and $a_\zeta^\dagger a_\eta^\dagger a_\iota a_\kappa$, and the electronic Hamiltonian $\hat{H}_e$, Eq. (7), in terms of the matrices $\mathbf{I}_{\mu\alpha}$, $\mathbf{X}_{\mu\alpha}$, $\mathbf{Y}_{\mu\alpha}$ and $\mathbf{Z}_{\mu\alpha}$ of all the qubits (cf. Sect. 6). The first three operators in Eq. (25) span (generates) the SU(2) Lie algebra (group), and $\hat{N}_{\mu\alpha}$ and $\hat{J}_{\mu\alpha}^2 = \hat{J}_{X\mu\alpha}^2 + \hat{J}_{Y\mu\alpha}^2 + \hat{J}_{Z\mu\alpha}^2$ are the Casimir operators of U(2) $\supset$ SU(2). The mutually commuting $\hat{J}_{Z\mu\alpha}$, $\hat{J}_{\mu\alpha}^2$ and $\hat{N}_{\mu\alpha}$ act on the spin-orbital $|\psi_\alpha\rangle$ and $|\psi_\mu\rangle$ as:

$$\hat{J}_{Z\mu\alpha}|\psi_\alpha\rangle = +\frac{1}{2}|\psi_\alpha\rangle; \quad \hat{J}_{Z\mu\alpha}|\psi_\mu\rangle = -\frac{1}{2}|\psi_\mu\rangle; \quad \hat{J}_{\mu\alpha}^2|\psi_{\alpha/\mu}\rangle = \frac{3}{4}|\psi_{\alpha/\mu}\rangle; \quad \hat{N}_{\mu\alpha}|\psi_{\alpha/\mu}\rangle = +1|\psi_{\alpha/\mu}\rangle; \quad (26)$$

with equivalent expressions in terms of the states (vectors) $|0\rangle_\alpha$ and $|1\rangle_\mu$ and matrices $\mathbf{I}_{\mu\alpha}$, $\mathbf{X}_{\mu\alpha}$, $\mathbf{Y}_{\mu\alpha}$, and $\mathbf{Z}_{\mu\alpha}$.

From this point to the end of this section, we will concentrate on the END/QC expressions for QC programming. Therefore, we will prioritize the matrix-vector representation of END/QC [in terms of the matrices $\mathbf{I}_{\mu\alpha}$, $\mathbf{X}_{\mu\alpha}$, $\mathbf{Y}_{\mu\alpha}$, and $\mathbf{Z}_{\mu\alpha}$ acting on the computational basis states (vectors) $|0\rangle_\alpha$ and $|1\rangle_\mu$ ] over its equivalent operator-wavefunction representation (in terms of the operators $\hat{N}_{\mu\alpha}$, $\hat{J}_{X\mu\alpha}$, $\hat{J}_{Y\mu\alpha}$ and $\hat{J}_{Z\mu\alpha}$ acting on the spin-orbitals $|\psi_\alpha\rangle$ and $|\psi_\mu\rangle$]. Nevertheless, the equivalences between both representations can be obtained straightforwardly [cf. Eq. (25)].

$\mathbf{u}(\lambda_{\mu\alpha}) \in U(2)$ can be factorized in terms of a global phase angle $\phi_0 \in \mathbb{R}$ and three extrinsic Euler angles $\phi_1$, $\phi_2$ and $\phi_3 \in \mathbb{R}$ in the z-y-z axis convention: $\mathbf{u}(\lambda_{\mu\alpha}) = \exp(i\phi_0) \mathbf{R}_z(\phi_1) \mathbf{R}_y(\phi_2) \mathbf{R}_z(\phi_3)$ [14], where $\mathbf{R}_i(\phi_j)$ is the 2 x 2 rotation matrix around the axis $i$ by an angle $\phi_j$. Then, from Eq. (23), one obtains $\phi_0 = 0$, $\phi_1 = \omega_{\mu\alpha}$, $\phi_2 = 2\rho_{\mu\alpha}$, and $\phi_3 = -\omega_{\mu\alpha}$ so that

$$\mathbf{u}(\lambda_{\mu\alpha}) = \mathbf{R}(\rho_{\mu\alpha}, \omega_{\mu\alpha}) = \mathbf{R}_z(+\omega_{\mu\alpha}) \mathbf{R}_y(2\rho_{\mu\alpha}) \mathbf{R}_z(-\omega_{\mu\alpha})$$
$$= \begin{bmatrix} e^{-i\frac{\omega_{\mu\alpha}}{2}} & 0 \\ 0 & e^{+i\frac{\omega_{\mu\alpha}}{2}} \end{bmatrix} \begin{bmatrix} \cos(\rho_{\mu\alpha}) & -\sin(\rho_{\mu\alpha}) \\ \sin(\rho_{\mu\alpha}) & \cos(\rho_{\mu\alpha}) \end{bmatrix} \begin{bmatrix} e^{+i\frac{\omega_{\mu\alpha}}{2}} & 0 \\ 0 & e^{-i\frac{\omega_{\mu\alpha}}{2}} \end{bmatrix}. \quad (27)$$

This additional factorization per qubit is useful in subsequent QC implementations. Alternatively, $\mathbf{u}(\lambda_{\mu\alpha})$ can be expressed as a general (1-norm quaternion) rotation by an angle $\rho_{\mu\alpha} \geq 0$ about an axis with unit vector $\mathbf{n} = (n_x, n_y, n_z)$, $\mathbf{n}^2 = 1$, [14]:

$$\mathbf{u}(\lambda_{\mu\alpha}) = \mathbf{R}(\rho_{\mu\alpha}, \omega_{\mu\alpha}) = \cos(\rho_{\mu\alpha})\mathbf{I}_{\mu\alpha} - i\sin(\rho_{\mu\alpha})(n_x\mathbf{X}_{\mu\alpha} + n_y\mathbf{Y}_{\mu\alpha} + n_z\mathbf{Z}_{\mu\alpha})$$
$$= \cos(\rho_{\mu\alpha})\mathbf{I}_{\mu\alpha} - i\sin(\rho_{\mu\alpha})\left[\sin(-\omega_{\mu\alpha})\mathbf{X}_{\mu\alpha} + \cos(\omega_{\mu\alpha})\mathbf{Y}_{\mu\alpha}\right]; \quad (28)$$

from which $n_x = \sin(-\omega_{\mu\alpha})$, $n_y = \cos(\omega_{\mu\alpha})$, and $n_z = 0$; then, $\mathbf{u}(\lambda_{\mu\alpha}) = \mathbf{R}(\rho_{\mu\alpha}, \omega_{\mu\alpha})$ is a rotation by an angle $\rho_{\mu\alpha}$ about an axis on the x-y plane and tilted by an angle $\omega_{\mu\alpha}$ from the y-axis. For constant angles $\omega_{\mu\alpha}$, the matrices $\{\mathbf{u}(\lambda_{\mu\alpha}) = \mathbf{R}(\rho_{\mu\alpha}, \omega_{\mu\alpha} = \text{constant})\}$ form a one-real-parameter continuous subgroup $\subset$ SU(2). Since the SU(2) matrices are identical to the matrices of the SU(2) irreducible representation $D^{J=1/2}$, the tensor product in Eq. (21) goes along with the products of the SU(2) irreducible representations $D^{J=1/2}$, which in general produce reducible representations, e.g. $D^{J=1/2} \otimes D^{J=1/2} = D^{J=0} \oplus D^{J=1}$ [42].

To obtain the components of the END/QC equations of motion for the model systems in QC form, we should first express the corresponding wavefunction and Hamiltonian in that form. The reference Slater determinant state $|\overline{0}\rangle$ is now [cf. Eq. (6)]

$$|\overline{0}\rangle = |0_\alpha 0_\beta \ldots 0_{N_e}\rangle = \otimes_\alpha^{N_e} |0_\alpha\rangle = |0_\alpha\rangle \otimes |0_\beta\rangle \otimes \ldots |0_{N_e}\rangle. \tag{29}$$

Then, through Eqs. (21)-(27), the END/QC trial wavefunction $|\Psi(\lambda)\rangle$, Eq. (18), corresponds to the QC state $|\Psi(\rho, \omega)\rangle$ with real parameters $\rho = (\rho_{\mu\alpha})$ and $\omega = (\omega_{\mu\alpha})$

$$|\Psi(\rho, \omega)\rangle = \mathbf{R}_{Total}|\overline{0}\rangle = \otimes_{(\mu\alpha)}^{N_e} \mathbf{R}(\rho_{\mu\alpha}, \omega_{\mu\alpha})|\overline{0}\rangle = \otimes_{(\mu\alpha)}^{N_e} \mathbf{R}_z(+\omega_{\mu\alpha})\mathbf{R}_y(2\rho_{\mu\alpha})\mathbf{R}_z(-\omega_{\mu\alpha})|\overline{0}\rangle$$
$$= \otimes_\alpha^{N_e} |\phi_\alpha\rangle; \qquad |\phi_\alpha\rangle = \cos(\rho_{\mu\alpha})|0_\alpha\rangle + \exp(+i\omega_{\mu\alpha})\sin(\rho_{\mu\alpha})|1_\mu\rangle; \tag{30}$$

where the original multiple-qubit states $|\phi_\alpha\rangle$ in Eq. (16) simplifies into standard single-qubit Bloch sphere states $|\phi_\alpha\rangle$ in Eq. (30). In addition, by mapping the operators $a_\eta^\dagger a_\zeta$ and $a_\zeta^\dagger a_\eta^\dagger a_\iota a_\kappa$ into the identity and Pauli matrices via Eq. (25) [e.g., $a_\alpha^\dagger a_\alpha \leftrightarrow (\mathbf{I}_{\mu\alpha} + \mathbf{Z}_{\mu\alpha})/2$, etc.], we can encode the electronic Hamiltonian $\hat{H}_e$, Eq. (7), as

$$\hat{H}_e = \sum_{i=1}^{N_h} \tilde{h}_i \mathbf{h}_i \tag{31}$$

where the coefficients $\tilde{h}_i$ are combinations of the original one-electron $h_{\zeta\eta}$ and two-electron $\langle\zeta\eta|\kappa\iota\rangle$ integrals in Eq.(7), and the unitary matrices $\mathbf{h}_i$ are combinations of the identity and Pauli matrices. The explicit expressions of the $\tilde{h}_i$ and $\mathbf{h}_i$ depend on the chemical features of the model system under consideration. We will present examples of these expressions for one-electron diatomic molecules in the following section and for additional molecules in a sequel. From Eq.

(30), we can obtain the derivatives of the END/QC trial wavefunction $|\Psi(\boldsymbol{\rho}, \boldsymbol{\omega})\rangle$ with respect to its variational parameters $\rho_k$ and $\omega_k$ as

$$\frac{\partial |\Psi(\boldsymbol{\rho}, \boldsymbol{\omega})\rangle}{\partial \rho_k} = \sum_{j=1}^{N_e} f_{k,j}^{\rho} \mathbf{R}_{k,j}^{\rho} |\overline{0}\rangle;$$
$$\mathbf{R}_{k,j}^{\rho} = \mathbf{R}(\rho_1, \omega_1) \otimes \ldots \mathbf{R}_z(+\omega_j) \mathbf{R}_y(2\rho_j) \mathbf{Y}_j \mathbf{R}_z(-\omega_j) \otimes \ldots \mathbf{R}(\rho_{N_e}, \omega_{N_e}) |\overline{0}\rangle \quad (32)$$

and

$$\frac{\partial |\Psi(\boldsymbol{\rho}, \boldsymbol{\omega})\rangle}{\partial \omega_k} = \sum_{j=1}^{N_e} f_{k,j}^{\omega} \mathbf{R}_{k,j}^{\omega} |\overline{0}\rangle = \sum_{j=1}^{N_e} \left( f_{k,j}^{+\omega} \mathbf{R}_{k,j}^{+\omega} + f_{k,j}^{-\omega} \mathbf{R}_{k,j}^{-\omega} \right) |\overline{0}\rangle;$$
$$\mathbf{R}_{k,j}^{\omega} = \mathbf{R}_{k,j}^{+\omega} - \mathbf{R}_{k,j}^{-\omega};$$
$$\mathbf{R}_{k,j}^{+\omega} = \mathbf{R}(\rho_1, \omega_1) \otimes \ldots \mathbf{R}_z(+\omega_j) \mathbf{Z}_j \mathbf{R}_y(2\rho_j) \mathbf{R}_z(-\omega_j) \otimes \ldots \mathbf{R}(\rho_{N_e}, \omega_{N_e}); \quad (33)$$
$$\mathbf{R}_{k,j}^{-\omega} = \mathbf{R}(\rho_1, \omega_1) \otimes \ldots \mathbf{R}_z(+\omega_j) \mathbf{R}_y(2\rho_j) \mathbf{R}_z(-\omega_j) \mathbf{Z}_j \otimes \ldots \mathbf{R}(\rho_{N_e}, \omega_{N_e});$$

where $j, k = (\mu\alpha) = 1, 2 \ldots N_e$, $\mathbf{R}_{k,j}^{\rho}$, $\mathbf{R}_{k,j}^{\omega}$, $\mathbf{R}_{k,j}^{+\omega}$, and $\mathbf{R}_{k,j}^{-\omega}$ are defined, and their associated factors are:

$$f_{k,j}^{\rho} = -\delta_{kj} i; \quad f_{k,j}^{\pm\omega} = \mp\delta_{k,j}\left(\frac{i}{2}\right); \quad f_{k,j}^{\omega} = f_{k,j}^{+\omega} = -f_{k,j}^{-\omega}. \quad (34)$$

Our Eqs. (32) and (33) are equivalent but not identical to their VQS counterparts in Ref. [17]; the main differences between those and our expressions lie in the occurrence of two derivative terms per each angle $\omega_k$, and in the content of some terms. Differences aside, our equations admit a QC implementation similar to that in Ref.[17], cf. next section. From Eqs. (31)-(34), we can obtain the metric matrix $\mathbf{M} = (M_{pq})$ and the energy gradient vector $\mathbf{V} = (V_p)$ with respect to the variational parameters $\{\xi_p\}$ and $\{\xi_p'\} = \{\rho_k, \omega_k\}$ [cf. Eq. (4)]:

$$M_{pq} = i\frac{\partial \langle \Psi(\xi)|}{\partial \xi_p}\frac{\partial |\Psi(\xi')\rangle}{\partial \xi'_q} + \text{H. c.} = \sum_{j,k=1}^{N_e,N_e}\left(if_{p,j}^{\xi*}f_{q,k}^{\xi'}\langle \overline{0}|\mathbf{R}_{p,j}^{\xi\dagger}\mathbf{R}_{q,k}^{\xi'}|\overline{0}\rangle + \text{H. c.}\right); \tag{35}$$

and

$$V_p = \frac{\partial \langle \Psi(\xi)|}{\partial \xi_p}\hat{H}_e|\Psi(\xi)\rangle + \text{H. c.} = \sum_{j,k=1,1}^{N_e,N_h}\left(f_{p,j}^{\xi*}\tilde{h}_k\langle \overline{0}|\mathbf{R}_{p,j}^{\xi\dagger}\mathbf{h}_k\mathbf{R}_{Total}|\overline{0}\rangle + \text{H. c.}\right); \tag{36}$$

where $\{f_{p,j}^{\xi}\}$ and $\{f_{p,j}^{\xi'}\} = \{f_{p,j}^{\rho}, f_{p,j}^{\omega}\}$, and $\{\mathbf{R}_{p,j}^{\xi}\}$ and $\{\mathbf{R}_{p,j}^{\xi'}\} = \{\mathbf{R}_{p,j}^{\rho}, \mathbf{R}_{p,j}^{\omega}\}$, cf. Eqs. (32)-(34). Each summand between parentheses in Eqs. (35) and (36) contains one, two, or four terms of the form (cf. Ref. [17]):

$$a\,\text{Re}\left[\exp(i\alpha)\langle \overline{0}|\tilde{\mathbf{U}}|\overline{0}\rangle\right] \tag{37}$$

where the QC parameters $a$ and $\alpha \in \mathbb{R}$, and the unitary matrix $\tilde{\mathbf{U}}$ are:

$$\begin{aligned}
&\text{For terms of } M_{pq}: \quad a = 2\left|if_{p,j}^{\xi*}f_{q,k}^{\xi'}\right|; \quad \alpha = \arg\left(if_{p,j}^{\xi*}f_{q,k}^{\xi'}\right); \quad \tilde{\mathbf{U}} = \mathbf{R}_{p,j}^{\xi\dagger}\mathbf{R}_{q,k}^{\xi'}; \\
&\text{For terms of } V_p: \quad a = 2\left|f_{p,j}^{\xi*}\tilde{h}_k\right|; \quad \alpha = \arg\left(f_{p,j}^{\xi*}\tilde{h}_k\right); \quad \tilde{\mathbf{U}} = \mathbf{R}_{p,j}^{\xi\dagger}\mathbf{h}_k\mathbf{R}_{Total};
\end{aligned} \tag{38}$$

where now $\{f_{p,j}^{\xi}\}$ and $\{f_{p,j}^{\xi'}\} = \{f_{p,j}^{\rho}, f_{p,j}^{+\omega}, f_{p,j}^{-\omega}\}$, and $\{\mathbf{R}_{p,j}^{\xi}\}$ and $\{\mathbf{R}_{p,j}^{\xi'}\} = \{\mathbf{R}_{p,j}^{\rho}, \mathbf{R}_{p,j}^{+\omega}, \mathbf{R}_{p,j}^{-\omega}\}$, cf. Eqs. (32)-(34). The terms $\text{Re}\left[\exp(i\alpha)\langle \overline{0}|\tilde{\mathbf{U}}|\overline{0}\rangle\right]$ in Eq. (37) can be evaluated with our END/QC versions of the VQS circuit in Ref.[17]. We will illustrate such QC evaluations in the following section.

## 6. Application of END/QC/VQS to One-Electron Diatomic Molecules.

As a *proof of concept* and for illustration's sake, we will now apply END/QC/VQS to simulate the pure electronic dynamics of one-electron diatomic molecules. We will consider the general case of hetero-nuclear diatomic molecules that contains the homo-nuclear ones as a simple subcase. In the framework of the model systems defined in Sect. 4, each of these molecules forms a single one-electron unit furnished with a minimal basis set of two Slater-type atomic orbitals[36] $\{\varphi_A, \varphi_B\}$, where $A$ and $B$ are the labels of the nuclei. With this basis set, we can construct one occupied $\psi_\alpha$ and one unoccupied $\psi_\mu$ spin-orbital:

$$\begin{aligned}\psi_\alpha(\mathbf{x}_1) &= \tilde{\psi}_\alpha(\mathbf{r}_1)\sigma(s_1) = \left[c_{A\alpha}\varphi_A(\mathbf{r}_1) + c_{B\alpha}\varphi_B(\mathbf{r}_1)\right]\sigma(s_1) \quad \text{occupied (hole);} \\ \psi_\mu(\mathbf{x}_1) &= \tilde{\psi}_\mu(\mathbf{r}_1)\sigma(s_1) = \left[c_{A\mu}\varphi_A(\mathbf{r}_1) + c_{B\mu}\varphi_B(\mathbf{r}_1)\right]\sigma(s_1) \quad \text{unoccupied (particle);}\end{aligned} \quad (39)$$

where $\tilde{\psi}_\alpha(\mathbf{r}_1)$ and $\tilde{\psi}_\mu(\mathbf{r}_1)$ are the highest occupied and lowest unoccupied molecular orbitals (HOMO and LUMO), respectively, $\sigma(s_1)$ is a one-electron spin eigenfunction, and $c_{A\alpha}$, $c_{B\alpha}$, $c_{A\mu}$ and $c_{B\mu}$ are the molecular orbital coefficients from a self-consistent-field HF calculation[36]. The transformed occupied $\phi_\alpha$ and unoccupied $\phi_\mu$ spin-orbitals and the END/QC total trial wavefunction $|\Psi(\rho_{\mu\alpha},\omega_{\mu\alpha})\rangle$ from the reference state $|\bar{0}\rangle = |\psi_\alpha\rangle$ are [cf. Eqs. (29)-(30)]

$$\begin{aligned}\phi_\alpha &= \hat{U}\left[\mathbf{R}(\rho_{\mu\alpha},\omega_{\mu\alpha})\right]\psi_\alpha = \cos(\rho_{\mu\alpha})\psi_\alpha + \exp(+i\omega_{\mu\alpha})\sin(\rho_{\mu\alpha})\psi_\mu; \\ \phi_\mu &= \hat{U}\left[\mathbf{R}(\rho_{\mu\alpha},\omega_{\mu\alpha})\right]\psi_\mu = -\exp(-i\omega_{\mu\alpha})\sin(\rho_{\mu\alpha})\psi_\alpha + \cos(\rho_{\mu\alpha})\psi_\mu; \\ |\Psi(\rho_{\mu\alpha},\omega_{\mu\alpha})\rangle &= \hat{U}\left[\mathbf{R}(\rho_{\mu\alpha},\omega_{\mu\alpha})\right]|\bar{0}\rangle = |\cos(\rho_{\mu\alpha})\psi_\alpha + \exp(+i\omega_{\mu\alpha})\sin(\rho_{\mu\alpha})\psi_\mu\rangle.\end{aligned} \quad (40)$$

The electronic Hamiltonian $\hat{H}_e$ of the considered molecules is [cf. Eq. (7)]

$$\hat{H}_e = h_{\alpha\alpha}a_\alpha^\dagger a_\alpha + h_{\mu\mu}a_\mu^\dagger a_\mu + h_{\mu\alpha}\left(a_\mu^\dagger a_\alpha + a_\alpha^\dagger a_\mu\right) \quad (41)$$

where the one-electron $h_{\zeta\eta}$ integrals $\in \mathbb{R}$ so that $h_{\mu\alpha} = h_{\alpha\mu}$. Notice that within the present atomic basis set, the Hamiltonian $\hat{H}_e$ is exact. $h_{\alpha\alpha}$ and $h_{\mu\mu}$ are the orbital energies of the HOMO $\tilde{\psi}_\alpha$ and LUMO $\tilde{\psi}_\mu$, respectively, and $\Delta_{\mu\alpha} = h_{\mu\mu} - h_{\alpha\alpha} > 0$ is the HOMO-LUMO energy gap. From Eqs. (40)-(41), we can obtain the energy of the considered systems as:

$$E(\rho_{\mu\alpha},\omega_{\mu\alpha}) = \langle \Psi(\rho_{\mu\alpha},\omega_{\mu\alpha})|\hat{H}_e|\Psi(\rho_{\mu\alpha},\omega_{\mu\alpha})\rangle$$
$$= \cos^2(\rho_{\mu\alpha})h_{\alpha\alpha} + \sin^2(\rho_{\mu\alpha})h_{\mu\mu} + \sin(2\rho_{\mu\alpha})\cos(\omega_{\mu\alpha})h_{\mu\alpha}. \quad (42)$$

To derive and implement the END/QC equations motion, we switch to the matrix-vector representation of END/QC delineated in Eqs. (29)-(38). Then, in the present case, the transformed occupied $|\phi_\alpha\rangle$ and unoccupied $|\phi_\mu\rangle$ computational basis states, and the END/QC total trial wavefunction $|\Psi(\xi_1,\xi_2,\xi_3)\rangle$ from the reference state $|\bar{0}\rangle = |0_\alpha\rangle$ are

$$|\Psi(\xi_1,\xi_2,\xi_3)\rangle = \mathbf{R}_{Total}|\bar{0}\rangle = \mathbf{R}_3(\xi_3)\mathbf{R}_2(\xi_2)\mathbf{R}_1(\xi_1)|\bar{0}\rangle = |\phi_\alpha\rangle;$$
$$|\phi_\alpha\rangle = \mathbf{R}_3(\xi_3)\mathbf{R}_2(\xi_2)\mathbf{R}_1(\xi_1)|0_\alpha\rangle = \cos(\rho_{\mu\alpha})|0_\alpha\rangle + \exp(+i\omega_{\mu\alpha})\sin(\rho_{\mu\alpha})|1_\mu\rangle; \quad (43)$$
$$|\phi_\mu\rangle = \mathbf{R}_3(\xi_3)\mathbf{R}_2(\xi_2)\mathbf{R}_1(\xi_1)|1_\mu\rangle = -\exp(-i\omega_{\mu\alpha})\sin(\rho_{\mu\alpha})|0_\alpha\rangle + \cos(\rho_{\mu\alpha})|1_\mu\rangle.$$

In some of the above expressions, we have adopted a simpler notation, more amenable for QC coding, where $\xi_1 = \omega_{\mu\alpha}$, $\xi_2 = \rho_{\mu\alpha}$, $\xi_3 = \omega'_{\mu\alpha}$, $\mathbf{R}_1(\xi_1) = \mathbf{R}_z(-\omega_{\mu\alpha})$, $\mathbf{R}_2(\xi_2) = \mathbf{R}_y(2\rho_{\mu\alpha})$, and $\mathbf{R}_3(\xi_3) = \mathbf{R}_z(+\omega'_{\mu\alpha})$, and $\omega_{\mu\alpha} \to \omega'_{\mu\alpha}$, cf. Eq. (27). From Eq. (43), the derivatives of $|\Psi(\xi_1,\xi_2,\xi_3)\rangle$ with respect to the variational parameters $\xi_1$, $\xi_2$, and $\xi_3$ are

$$\frac{\partial|\Psi(\xi_1,\xi_2,\xi_3)\rangle}{\partial\xi_1} = f_1\mathbf{R}^{\xi_1}|\bar{0}\rangle = f_1\mathbf{R}_3(\xi_3)\mathbf{R}_2(\xi_2)\mathbf{R}_1(\xi_1)\mathbf{Z}|\bar{0}\rangle;$$
$$\frac{\partial|\Psi(\xi_1,\xi_2,\xi_3)\rangle}{\partial\xi_2} = f_2\mathbf{R}^{\xi_2}|\bar{0}\rangle = f_2\mathbf{R}_3(\xi_3)\mathbf{R}_2(\xi_2)\mathbf{Y}\,\mathbf{R}_1(\xi_1)|\bar{0}\rangle; \quad (44)$$
$$\frac{\partial|\Psi(\xi_1,\xi_2,\xi_3)\rangle}{\partial\xi_3} = f_3\mathbf{R}^{\xi_3}|\bar{0}\rangle = f_3\mathbf{R}_3(\xi_3)\mathbf{Z}\mathbf{R}_2(\xi_2)\mathbf{R}_1(\xi_1)|\bar{0}\rangle;$$

where $f_1$, $f_2$ and $f_3$ are equivalent to $f_{1,1}^{-\omega}$, $f_{1,1}^{\rho}$ and $f_{1,1}^{+\omega}$ in Eq. (34), respectively,

$$f_1 = +\frac{i}{2};\quad f_2 = -i;\quad f_3 = -\frac{i}{2}. \quad (45)$$

Through Eq. (25), we can encode the electronic Hamiltonian $\hat{H}_e$ of Eq. (41) into the QC form of Eq. (31) as

$$\hat{H}_e = \tilde{h}_I \mathbf{I} + \tilde{h}_X \mathbf{X} + \tilde{h}_Y \mathbf{Y} + \tilde{h}_Z \mathbf{Z};$$
$$\tilde{h}_I = \frac{1}{2}(h_{\alpha\alpha} + h_{\mu\mu}); \ \tilde{h}_X = h_{\mu\alpha}; \ \tilde{h}_Y = 0; \ \tilde{h}_Z = \frac{1}{2}(h_{\alpha\alpha} - h_{\mu\mu}). \quad (46)$$

From Eqs. (44)-(46), we obtain the components of the END/QC equations of motion as

$$M_{\rho_{\mu\alpha},\rho_{\mu\alpha}} = 0; \ M_{\omega_{\mu\alpha},\omega_{\mu\alpha}} = 0; \ M_{\rho_{\mu\alpha},\omega_{\mu\alpha}} = \tilde{M}_{21} + \tilde{M}_{23} = -\sin(2\rho_{\mu\alpha});$$
$$\tilde{M}_{21} = if_2^* f_1 \langle \bar{0} | \mathbf{R}^{\xi_2\dagger} \mathbf{R}^{\xi_1} | \bar{0} \rangle + \text{H. c.}; \ \tilde{M}_{23} = if_2^* f_3 \langle \bar{0} | \mathbf{R}^{\xi_2\dagger} \mathbf{R}^{\xi_3} | \bar{0} \rangle + \text{H. c.}; \quad (47)$$

and

$$V_{\rho_{\mu\alpha}} = \sum_{k=I,X,Y,Z} \tilde{V}_{2k} = -2\tilde{h}_Z \sin(2\rho_{\mu\alpha}) + 2\tilde{h}_X \cos(2\rho_{\mu\alpha})\cos(\omega_{\mu\alpha});$$
$$V_{\omega_{\mu\alpha}} = \sum_{k=I,X,Y,Z} (\tilde{V}_{1k} + \tilde{V}_{3k}) = -\tilde{h}_X \sin(2\rho_{\mu\alpha})\sin(\omega_{\mu\alpha}); \quad (48)$$
$$\tilde{V}_{jk} = f_j^* \tilde{h}_k \langle \bar{0} | \mathbf{R}^{\xi_j\dagger} \hat{h}_k \mathbf{R}_{Total} | \bar{0} \rangle + \text{H. c.}; \ j = 1, 2, 3; \ k = \mathbf{I}, \mathbf{X}, \mathbf{Y}, \mathbf{Z}.$$

For non-null values in Eqs. (47) and (48), we write first the QC expressions of the elements of **M** and **V** obtained from Eqs. (44)-(46), and second their equivalent analytical expressions obtained from Eqs. (40)-(42). In terms of the latter, the END/QC equations of motion are

$$\begin{pmatrix} 0 & -\sin(2\rho_{\mu\alpha}) \\ +\sin(2\rho_{\mu\alpha}) & 0 \end{pmatrix} \begin{pmatrix} \dot{\rho}_{\mu\alpha} \\ \dot{\omega}_{\mu\alpha} \end{pmatrix} = \begin{pmatrix} -2\tilde{h}_Z \sin(2\rho_{\mu\alpha}) + 2\tilde{h}_X \cos(2\rho_{\mu\alpha})\cos(\omega_{\mu\alpha}) \\ -\tilde{h}_X \sin(2\rho_{\mu\alpha})\sin(\omega_{\mu\alpha}) \end{pmatrix}. \quad (49)$$

To obtain **M** and **V** for the END/QC/VQS simulations, we need to evaluate their basic elements $\tilde{M}_{jk}$ and $\tilde{V}_{jk}$ in Eqs. (47) and (48) in a QC fashion. After recasting these elements in the form of Eq. (37), we can evaluate them with the five quantum circuits shown in Figs. 2-6; these circuits are the END/QC versions of the VQS circuit in Fig. 2 of Ref.[17]. Each of the present circuits has one ancillary qubit prepared in the state $(|0\rangle + e^{i\alpha}|1\rangle)/\sqrt{2}$, where the values of the circuit parameter $\alpha$ and $a$, Eq. (37), are obtained with Eq. (38). The values of $\alpha$ and $a$ employed in the present QC evaluations are listed in Table 1. All the ancillary qubits have the same operators: two **X** gates and two control gates, but in different orders and with varying targets. Each of the present circuits has also one register qubit initially prepared in the END/QC reference state $|\bar{0}\rangle$. The operators in the register qubits correspond to those involved in each evaluated element $\tilde{M}_{jk}$ and $\tilde{V}_{jk}$, cf. Eq. (47)-(48). All these circuits end with a measurement of the ancillary qubit in the $\{|+\rangle, |-\rangle\}$ basis[14], where $|\pm\rangle = (|0\rangle \pm |1\rangle)/\sqrt{2}$ are the eigenvectors of the Pauli matrix **X** with eigenvalues +1 and -1, respectively[14]. The average measurement provides the expectation value $\langle \mathbf{X} \rangle = \mathrm{Re}\left(e^{i\alpha}\langle\bar{0}|\tilde{U}|\bar{0}\rangle\right)$, cf. Eqs. (37)-(38), as explained in the following paragraph.

All the present quantum circuits operate similarly and we will elucidate their execution by analyzing the sequential operations in the circuit for $\tilde{M}_{23}$ shown in Fig. 3. At each of the steps drawn in Fig. 3, the circuit total state $|\Psi_i\rangle$ successively is

$$|\Psi_0\rangle = \frac{1}{\sqrt{2}}\left(|0\rangle + e^{i\alpha}|1\rangle\right) \otimes |\bar{0}\rangle = \frac{1}{\sqrt{2}}\left(|0\rangle \otimes |\bar{0}\rangle + e^{i\alpha}|1\rangle \otimes |\bar{0}\rangle\right);$$

$$|\Psi_1\rangle = \frac{1}{\sqrt{2}}\left(|1\rangle \otimes \mathbf{R}_1|\bar{0}\rangle + e^{i\alpha}|0\rangle \otimes \mathbf{R}_1|\bar{0}\rangle\right);$$

$$|\Psi_2\rangle = \frac{1}{\sqrt{2}}\left(|1\rangle \otimes \mathbf{YR}_1|\bar{0}\rangle + e^{i\alpha}|0\rangle \otimes \mathbf{R}_1|\bar{0}\rangle\right);$$

$$|\Psi_3\rangle = \frac{1}{\sqrt{2}}\left(|0\rangle \otimes \mathbf{R}_2\mathbf{YR}_1|\bar{0}\rangle + e^{i\alpha}|1\rangle \otimes \mathbf{R}_2\mathbf{R}_1|\bar{0}\rangle\right);$$

$$|\Psi_4\rangle = \frac{1}{\sqrt{2}}\left(|0\rangle \otimes \mathbf{R}_2\mathbf{YR}_1|\bar{0}\rangle + e^{i\alpha}|1\rangle \otimes \mathbf{ZR}_2\mathbf{R}_1|\bar{0}\rangle\right);$$

$$|\Psi_5\rangle = \frac{1}{\sqrt{2}}\left(|0\rangle \otimes \mathbf{R}_3\mathbf{R}_2\mathbf{YR}_1|\bar{0}\rangle + e^{i\alpha}|1\rangle \otimes \mathbf{R}_3\mathbf{ZR}_2\mathbf{R}_1|\bar{0}\rangle\right);$$

(50)

By changing from the $\{|0\rangle, |1\rangle\}$ basis to the $\{|+\rangle, |-\rangle\}$ one in $|\Psi_5\rangle$, we obtain

$$|\Psi_5\rangle = \frac{1}{2}\left[|+\rangle \otimes (\mathbf{R}_3\mathbf{R}_2\mathbf{YR}_1 + e^{i\alpha}\mathbf{R}_3\mathbf{ZR}_2\mathbf{R}_1)|\overline{0}\rangle + |-\rangle \otimes (\mathbf{R}_3\mathbf{R}_2\mathbf{YR}_1 - e^{i\alpha}\mathbf{R}_3\mathbf{ZR}_2\mathbf{R}_1)|\overline{0}\rangle\right];$$
$$= \frac{1}{2}\left[|+\rangle \otimes (\mathbf{R}^{\xi_2} + e^{i\alpha}\mathbf{R}^{\xi_3})|\overline{0}\rangle + |-\rangle \otimes (\mathbf{R}^{\xi_2} - e^{i\alpha}\mathbf{R}^{\xi_3})|\overline{0}\rangle\right]; \quad (51)$$

where the unitary matrices $\mathbf{R}^{\xi_2}$ and $\mathbf{R}^{\xi_3}$ are defined in Eq. (44). Then, from Eq. (51), the probability $P(|\pm\rangle)$ to find the ancillary qubit in the state $|\pm\rangle$ is

$$P(|\pm\rangle) = \frac{1}{4}\langle\overline{0}|(\mathbf{R}^{\xi_2} \pm e^{i\alpha}\mathbf{R}^{\xi_3})^\dagger (\mathbf{R}^{\xi_2} \pm e^{i\alpha}\mathbf{R}^{\xi_3})|\overline{0}\rangle$$
$$= \frac{1}{2} \pm \frac{1}{2}\mathrm{Re}\left(e^{i\alpha}\langle\overline{0}|\mathbf{R}^{\xi_2\dagger}\mathbf{R}^{\xi_3}|\overline{0}\rangle\right) = \frac{1}{2} \pm \frac{1}{2}\mathrm{Re}\left(e^{i\alpha}\langle\overline{0}|\tilde{\mathbf{U}}|\overline{0}\rangle\right); \quad (52)$$

where $\tilde{\mathbf{U}} = \mathbf{R}^{\xi_2\dagger}\mathbf{R}^{\xi_3}$ is another unitary matrix. Then, the average measurement of the ancillary qubit in the $\{|+\rangle, |-\rangle\}$ basis is the expectation value $\langle\mathbf{X}\rangle$

$$\langle\mathbf{X}\rangle = \langle\Psi_5|\mathbf{X}\otimes\mathbf{I}|\Psi_5\rangle = (+1)P(|+\rangle) + (-1)P(|-\rangle) = \mathrm{Re}\left(e^{i\alpha}\langle\overline{0}|\tilde{\mathbf{U}}|\overline{0}\rangle\right); \quad (53)$$

which is the final result from the quantum circuit. From that, $\tilde{M}_{23}$ is (cf. Table 1)

$$\tilde{M}_{23} = if_2^* f_3 \langle\overline{0}|\mathbf{R}^{\xi_2\dagger}\mathbf{R}^{\xi_3}|\overline{0}\rangle + \mathrm{H.\,c.} = a\,\mathrm{Re}\left(e^{i\alpha}\langle\overline{0}|\tilde{\mathbf{U}}|\overline{0}\rangle\right) = -\sin(\rho_{\mu\alpha});$$
$$a = 2|if_2^* f_3| = 1; \quad \alpha = \arg(if_2^* f_3) = \frac{\pi}{2}; \quad \tilde{\mathbf{U}} = \mathbf{R}^{\xi_2\dagger}\mathbf{R}^{\xi_3}; \quad (54)$$

The remaining circuits operate in a similar way.

To appraise the accuracy and precision of these circuits, we will examine their results for the elements $\tilde{M}_{23}$, $\tilde{V}_{2,X}$, $\tilde{V}_{2,Z}$, and $\tilde{V}_{3,X}$ corresponding to the values of the END/QC variational parameters $\rho_{\mu\alpha}$ and $\omega_{\mu\alpha}$ listed in Table 1. These are the only elements having circuit-evaluated components $\mathrm{Re}\left(e^{i\alpha}\langle\overline{0}|\tilde{\mathbf{U}}|\overline{0}\rangle\right)$ not identical to zero $\forall\,\rho_{\mu\alpha}$ and $\omega_{\mu\alpha}$. We performed all these

circuit calculations on the QC software development kit QISKIT[43]. To appraise the accuracy of these QC calculations, we will consider the absolute error (AE), $\text{AE}\left[T^{QC}(n_s,i)\right]$, and mean AE (MAE), $\text{MAE}\left[T^{QC}(n_s)\right]$:

$$\text{AE}\left[T^{QC}(n_s,i)\right] = \left|T^{QC}(n_s,i) - T^{Exact}\right|; \quad \text{MAE}\left[T^{QC}(n_s)\right] = \frac{1}{N_r}\sum_{i=1}^{N_r}\text{AE}\left[T^{QC}(n_s,i)\right]; \quad (55)$$

where $T^{QC}(n_s,i)$ is the value of $\tilde{M}_{23}$, $\tilde{V}_{2,X}$, $\tilde{V}_{2,Z}$, or $\tilde{V}_{3,X}$ from the $i$ repetition of a QC calculation with $n_s$ shots, and $T^{Exact}$ is the value of the same element from its analytical expression, Eqs. (47)-(48). The $\text{MAE}\left[T^{QC}(n_s)\right]$ is the average of the $\text{AE}\left[T^{QC}(n_s,i)\right]$ over the total number of repetitions $N_r$. In Figs. 7-10, we plot the $\log_2\left\{\text{AE}\left[T^{QC}(n_s,i)\right]\right\}$ for $i$ =1 to $N_r$ = 1,000 repetitions, and the $\log_2\left\{\text{MAE}\left[T^{QC}(n_s)\right]\right\}$ vs. $\log_2(n_s)$. In each figure, for a given number of shots $n_s$, the individual values of $\log_2\left\{\text{AE}\left[T^{QC}(n_s,i)\right]\right\}$ appear as vertically scattered blue points, and the values of $\log_2\left\{\text{MAE}\left[T^{QC}(n_s)\right]\right\}$ appear as red start points/lines. In addition, in each figure, we plot the regression line corresponding to $\log_2\left\{\text{MAE}\left[T^{QC}(n_s)\right]\right\}$ vs. $\log_2(n_s)$ in black and report its slope $\tilde{\alpha}$, intercept $\tilde{\beta}$, and coefficient of determination $R^2$ in each figure caption. Remarkably, all the regression lines exhibit the same slope $\tilde{\alpha} \approx -1/2$ with a perfect correlation with $R^2 \approx 1$:

$$\log_2\left\{\text{MAE}\left[T^{QC}(n_s)\right]\right\} = \tilde{\alpha}\log_2(n_s) + \tilde{\beta} \approx -\frac{1}{2}\log_2(n_s) + \tilde{\beta};$$
$$\Rightarrow \text{MAE}\left[T^{QC}(n_s)\right] \approx 2^{\tilde{\beta}} n_s^{-1/2}; \quad (56)$$

*i.e.*, the asymptotic behavior of these errors with respect to $n_s$ is of the order $O\left(n_s^{-1/2}\right)$. To appraise the precision of these QC evaluations, we will now consider the standard deviation (SD), $\sigma(n_s)$, of the individual $\text{AE}\left[T^{QC}(n_s,i)\right]$ with respect to its $\text{MAE}\left[T^{QC}(n_s)\right]$:

$$\sigma(n_s) = \sqrt{\frac{1}{N_r}\sum_{i=1}^{N_r}\left\{\text{AE}\left[T^{QC}(n_s,i)\right] - \text{MAE}\left[T^{QC}(n_s)\right]\right\}^2} \; ; \tag{57}$$

where all the terms have been defined in/after Eq. (55). In Fig. 11, we plot the $\log_2\left[\sigma(n_s)\right]$ vs. $\log_2(n_s)$ of the QC calculations of $\tilde{M}_{23}$, $\tilde{V}_{2,X}$, $\tilde{V}_{2,Z}$, and $\tilde{V}_{3,X}$ and their corresponding regression lines. Like in the MAEs' case, all the regression lines of the SDs show the same slope $\tilde{\alpha} \approx -1/2$, with a perfect correlation with $R^2 \approx 1$. This demonstrates that the asymptotic behavior of the error spread with respect to $n_s$ is of the order $O\left(n_s^{-1/2}\right)$. The QC evaluations of the remaining elements $\tilde{M}_{jk}$ and $\tilde{V}_{jk}$ having circuit-evaluated components $\text{Re}\left(e^{i\alpha}\langle\bar{0}|\tilde{U}|\bar{0}\rangle\right)$ identical to zero $\forall\; \rho_{\mu\alpha}$ and $\omega_{\mu\alpha}$ exhibit same behaviors and trends in their accuracy and precision.

Finally, to elucidate a full END/QC/VQS simulation, we will examine its operations for the pure electronic dynamics of a $H_2^+$ molecule, an homonuclear diatomic system. Before examining computational aspects, we will briefly discuss the spatial symmetry of $H_2^+$ and its effect on the END/QC dynamics. $H_2^+$ has a $D_{\infty h}$ spatial symmetry, and its HOMO and LUMO, $\tilde{\psi}_\alpha$ and $\tilde{\psi}_\mu$, belong to the one-dimensional $D_{\infty h}$ irreducible representations of gerade $(\sigma_g)$ and ungerade $(\sigma_u)$ functions, respectively [36,44]. This discrepancy between representations dictates that $h_{\mu\alpha} = h_{\alpha\mu} = \tilde{h}_X = 0$ exactly [36,44], cf. Eq. (46). Due to this condition, the END/QC equations of motion for $H_2^+$, Eq. (49), can be solved *analytically* as

$$\begin{aligned}
\dot{\rho}_{\mu\alpha}(t) &= 0 \Rightarrow \rho_{\mu\alpha}(t) = \rho_{\mu\alpha}^0; \\
\dot{\omega}_{\mu\alpha}(t) &= -(h_{\mu\mu} - h_{\alpha\alpha}) \Rightarrow \omega_{\mu\alpha}(t) = -(h_{\mu\mu} - h_{\alpha\alpha})t + \omega_{\mu\alpha}^0; \\
\left| \Psi[\rho_{\mu\alpha}(t), \omega_{\mu\alpha}(t)] \right\rangle &= \left| \cos(\rho_{\mu\alpha}^0)\psi_\alpha + \exp[+i\omega_{\mu\alpha}(t)]\sin(\rho_{\mu\alpha}^0)\psi_\mu \right\rangle;
\end{aligned} \qquad (58)$$

where $\rho_{\mu\alpha}^0$ and $\omega_{\mu\alpha}^0$ are two initial conditions. Notice that the time-dependent exponential term of $\left| \Psi[\rho_{\mu\alpha}(t), \omega_{\mu\alpha}(t)] \right\rangle$ involves the angular frequency $(h_{\mu\mu} - h_{\alpha\alpha}) = \Delta_{\mu\alpha}$, *i.e.*, the HOMO-LUMO gap. $\left| \Psi[\rho_{\mu\alpha}(t), \omega_{\mu\alpha}(t)] \right\rangle$ remains stationary as an spin-orbital $\pm\psi_\alpha$ or $\pm\exp[+i\omega_{\mu\alpha}(t)]\psi_\mu \sim \pm\psi_\mu$ if $\rho_{\mu\alpha}^0 = k\pi$ or $=(2k-1)\pi/2$, $k=0, \pm1, \pm2 \ldots$, respectively. For values of $\rho_{\mu\alpha}^0$ different from the previous ones, $\left| \Psi[\rho_{\mu\alpha}(t), \omega_{\mu\alpha}(t)] \right\rangle$ superimposes both spin-orbitals $\psi_\alpha$ and $\psi_\mu$, undergoes a real dynamics, and leads to molecular properties evolving with angular frequency $\omega_{\mu\alpha} = (h_{\mu\mu} - h_{\alpha\alpha}) = \Delta_{\mu\alpha}$ (cf. the last paragraph of this section).

To perform an END/QC/VQS simulation according to the flowchart in Fig. 1, we will consider a $H_2^+$ molecule described with a minimal STO-3G basis set[36] (cf. Approximation 1, Sect. 4), and with a bond distance $R = 1.4$ a.u. The END/QC/VQS simulation of this system starts with task I that calculates the spin-orbital integrals $h_{\alpha\alpha} = -1.2528$ a.u. and $h_{\mu\mu} = -0.4756$ a.u., cf. Eq. (46), for the initial nuclear positions of $H_2^+$. This task is performed on a classical computer with the OED/ERD atomic integrals package [10] incorporated in our END code PACE[4]. In the present example, task I need not be repeated at each successive time step because the nuclear positions of $H_2^+$ do not change during pure electronic dynamics. Next, task II calculates the matrix **M** and vector **V** for the values of the END/QC variational parameters $\xi(t) = \rho_{\mu\alpha}(t)$ and $\omega_{\mu\alpha}(t)$ at each time step. This task is performed with the quantum circuits in Figs. 2-6, which calculate the basic elements $\tilde{M}_{jk}$ and $\tilde{V}_{jk}$ as discussed previously; these elements are subsequently combined to construct **M** and **V** via Eqs. (47)-(48). To appraise the accuracy and precision of the **M** and **V** calculations, we consider the root mean squared error (RMSE), $RMSE\left[\mathbf{A}^{QC}(n_s,i)\right]$ and the mean RMSE, $MRMSE\left[\mathbf{A}^{QC}(n_s)\right]$, for $\mathbf{A} = \mathbf{M}$ or $\mathbf{V}$ as

$$RMSE\left[\mathbf{A}^{QC}(n_s,i)\right] = +\sqrt{\frac{1}{\bar{N}}\sum_{j,k=1}^{\bar{N}}\left[A_{jk}^{QC}(n_s,i) - A_{jk}^{Exact}\right]^2};$$

$$MRMSE\left[\mathbf{A}^{QC}(n_s)\right] = \frac{1}{N_r}\sum_{i=1}^{N_r} RMSE\left[\mathbf{A}^{QC}(n_s,i)\right];$$

(59)

where $A_{jk}^{QC}(n_s,i)$ is the value of an element/component of **A** from the $i$ repetition of a QC calculation with $n_s$ shots, $A_{jk}^{Exact}$ is the value of the same element/component from an analytical expression, Eqs. (47)-(48), $\bar{N}$ is the number of elements/components of **A**, and $N_r$ is the total number of repetitions = 1,000; the RMSE is proportional to the Frobenius norm $\left\|\mathbf{A}^{QC} - \mathbf{A}^{Exact}\right\|$. Like in Figs. 7-10, we plot in Figs. 12-13 the $\log_2\left\{RMSE\left[\mathbf{A}^{QC}(n_s,i)\right]\right\}$, $\log_2\left\{MRMSE\left[\mathbf{A}^{QC}(n_s)\right]\right\}$, and the regression line of the latter vs. $\log_2(n_s)$ of $\mathbf{A} = \mathbf{M}$ and $\mathbf{V}$, respectively, for the values of the END/QC variational parameters $\rho_{\mu\alpha}$ and $\omega_{\mu\alpha}$ listed in Table 1. In addition, we plot in Fig. 11 the log₂ of the SD of the $RMSE\left[\mathbf{A}^{QC}(n_s,i)\right]$'s with respect to its $MRMSE\left[\mathbf{A}^{QC}(n_s)\right]$ and its regression line vs. $\log_2(n_s)$. Like in all previous QC calculations, the log₂-log₂ plots of the MRSMEs and SDs vs. $n_s$ show the same slopes $\tilde{\alpha} \approx -1/2$ with perfect linear correlations; furthermore, both metrics again scale asymptotically with an order $O(n_s^{-1/2})$. Finally, task III integrates the END/QC equations of motion with the current **M** and **V** over one time step $\Delta t$ to obtain the new END/QC variational parameters $\rho_{\mu\alpha}(t+\Delta t)$ and $\omega_{\mu\alpha}(t+\Delta t)$ for the subsequent time step. This task is performed with the differential equations solvers of PACE[4] on a classical computer. Alternatively, in the present example, the dynamics can be computed analytically with the expressions in Eq. (58).

To illustrate the dynamics of $H_2^+$ and gain chemical insight, we will present the time evolution of some molecular properties of $H_2^+$ from the time-dependent END/QC wavefunction $|\Psi[\rho_{\mu\alpha}(t), \omega_{\mu\alpha}(t)]\rangle$. A first property to consider is the one-electron density $\rho(\mathbf{r},t)$ [36]:

$$\rho(\mathbf{r}) = \left|\Psi[\rho_{\mu\alpha}(t), \omega_{\mu\alpha}(t)]\right|^2 = \sum_{A,B} P_{AB}(t)\varphi_A(\mathbf{r})\varphi_B^*(\mathbf{r}); \tag{60}$$

where $\mathbf{P}(t) = [P_{AB}(t)]$ is the one-electron density matrix in the atomic orbitals' basis $\{\varphi_A, \varphi_B\}$. $\rho(\mathbf{r},t)$ provides the probability density to find the electron of $H_2^+$ in position $\mathbf{r}$ at time $t$. The other properties are the electron Mulliken populations[36] on the nuclei $A$ and $B$, $N_A(t)$ and $N_B(t)$, respectively:

$$N_A(t) = 1 - N_B(t) = \sum_{C \in A} [\mathbf{P}(t)\mathbf{S}]_{CC}; \tag{61}$$

where $\mathbf{S} = (S_{AB})$ is the overlap matrix of the atomic orbitals $\{\varphi_A, \varphi_B\}$. $N_A(t)$ and $N_B(t)$ provide estimates of the total number of electrons around nuclei $A$ and $B$, respectively. In Fig. 14, we show six sequential snapshots of an END/QC/VQS/STO-3G computer animation of the pure electronic dynamics of $H_2^+$ at bond distance $R = 1.4$ a.u. and from the initial conditions $\rho_{\mu\alpha}^0 = 5^0$ and $\omega_{\mu\alpha}^0 = 0^0$. In each frame of Fig. 14, white spheres represent the fixed H nuclei, and the green cloud depicts an electron density $\rho(\mathbf{r},t)$ iso-value = 0.1 a.u. In the first frame at evolution time $t = 0$ fs, we notice that the initial condition $\rho_{\mu\alpha}^0 = 5^0$ distorts the ground-state density $\rho_{gs}(\mathbf{r}) = |\psi_\alpha(\mathbf{r})|^2$ conforming to the $D_{\infty h}$ molecular symmetry into an upright pear shape density $\rho(\mathbf{r},t=0)$ not conforming to $D_{\infty h}$ symmetry but to the lower $C_{\infty v}$ one. This reveals a spatial symmetry breaking[19,40,41] in $|\Psi[\rho_{\mu\alpha}(0), \omega_{\mu\alpha}(0)]\rangle$ and $\rho(\mathbf{r},t=0)$, which results from the combination of the spin-orbitals $\psi_\alpha$ and $\psi_\mu$ from different $D_{\infty h}$ irreducible representation in $|\Psi[\rho_{\mu\alpha}(0), \omega_{\mu\alpha}(0)]\rangle$ when $\rho_{\mu\alpha}^0 = 5^0$, cf. Eq. (58). In a sense, this superposition ("mixing") of the spin-orbitals and its resulting symmetry breaking propel the subsequent dynamics. In the following frames, $\rho(\mathbf{r},t)$ evolves to a $D_{\infty h}$-conforming shape at $t = 0.242$ fs, to a $C_{\infty v}$-conforming downright pear shape at $t = 0.484$ fs, again to a $D_{\infty h}$-conforming shape at $t = 0.726$ fs, and finally to the initial upright pear shape at $t = 0.968$ fs. This transformation of shapes repeats periodically in time. In chemical terms, these snapshots reveal that $H_2^+$ undergoes periodical intramolecular electron transfers between the two nuclei. Finally, in Fig. 15, we plot the electron Mulliken population on one of the nuclei of $H_2^+$ vs. time obtained by numerical integration of the END/QC equations, Eq. (49), with the Shampine-Gordon predict-evaluate-correct-evaluate method[45] in PACE[4], and obtained from the analytical solution in Eq. (58). Numerical and analytical results are identical as expected. Both types of Mulliken populations oscillate sinusoidally with the theoretical period $T = 2\pi/\omega_{\mu\alpha} = 2\pi/(h_{\mu\mu} - h_{\alpha\alpha}) = 2\pi/\Delta_{\mu\alpha} = 8.0844$ a.u., cf. Eq. (58), where $h_{\alpha\alpha} =$

-1.2528 a.u. and $h_{\mu\mu}$ = -0.4756 a.u. in $H_2^+$ with $R$ = 1.4 a.u.. Fig. 15 also reveals the aforesaid intramolecular electron transfers.

## 7. Conclusions and Future Work

In this manuscript, we present the first installment of the QC formulation of the END method[3,4] within the VQS scheme[17], i.e., END/QC/VQS. END[3,4] is a time-dependent, variational, direct, and non-adiabatic method to simulate various types of chemical reactions and scattering processes. As implemented in our code PACE[4] on *classical* computers, END adopts a total trial wavefunction that represents nuclei with frozen Gaussian wave packets and electrons with a single-determinantal state in the Thouless *non-unitary* representation[5]. The END equations of motions are obtained through the application of the TDVP[2] to the END trial wavefunction; this procedure renders a set of symplectic equations for the nuclear and electronic variational parameters $\xi$, $\mathbf{M}\dot{\xi} = \mathbf{V}$, where $\mathbf{M}$ is the phase-space metric matrix and $\mathbf{V}$ the energy gradient vector [2-4]. To implement END on *quantum* computers, we adopt the VQS scheme[17]: a *hybrid quantum/classical* approach to simulate symplectic equations of motion from the TDVP[2] and analogous variational principles[23-25]. In our case, an END/QC simulation in the VQS scheme involves three main tasks: **(I)** The calculation of atomic and molecular basis functions integrals on a *classical* computer, **(II)** the calculation of all the components of the END/QC equations of motion, $\mathbf{M}$ and $\mathbf{V}$, on a *quantum* computer, and **(III)** the time integration of those equations, $\mathbf{M}\dot{\xi} = \mathbf{V}$, on a *classical* computer. The philosophy of the VQS scheme[17] is to perform each numerical task on the type of computer, either *classical* or *quantum*, that delivers the most efficient result with current technology.

To derive the general END/QC formalism, we substitute Thouless *non-unitary* representation of the END single-determinantal electronic state[3,4] with Fukutome *unitary* representation[19] of the same state. Through this innovation, the new END/QC formalism fits directly into the unitary framework of QC[14]. Fukutome representation is based on the $U(K)$ Lie unitary group[19]; therefore, to understand the END/QC structure, we present a review of Fukutome representation that underlines its relationship with the $U(K)$ group and associated algebra via second quantization. In that context, we are the first to identify Fukutome unitary transformation matrices and transformed spin-orbitals as multi-qubit generalizations of one-qubit unitary matrices $U(2)$ and of one-qubit Bloch sphere states[14], respectively. In this investigation, we adopt a Fukutome unitary representation in terms of real parameters because it fits directly into the chemical systems under consideration. Accordingly, we also present a review of the particular form of the TDVP in terms of real variational parameters[2]. Nevertheless, it is possible to derive alternative formulations of END/QC in terms of the Fukutome unitary representation with complex parameters, or in terms of the Thouless non-unitary representation with complex parameters via LCU [20]; we will present these alternative approaches in a sequel soon.

The formulation and code implementation of END/QC for any type of dynamics and chemical systems are challenging endeavors. Therefore, in this investigation, we develop END/QC for pure electronic dynamics, i.e., for the time evolution of the electrons in the presence of fixed nuclei. Furthermore, within this dynamics, we derive END/QC for a family of model chemical systems defined by a set of four approximations; the latter are akin to semi-empirical approximations[22] employed in quantum chemistry[22]. In essence, these model systems consist of chemical units (atoms, molecules, monomers, etc.), each of them containing two effective electrons with opposite spins and represented with a minimal basis set. For these systems, Fukutome unitary transformation matrices factorize into one-qubit $U(2)$ matrices; in turn, each of these $U(2)$ matrices factorizes into three one-qubit rotational matrices $\mathbf{R}_z(+\omega_{\mu\alpha})$, $\mathbf{R}_y(2\rho_{\mu\alpha})$, and $\mathbf{R}_z(-\omega_{\mu\alpha})$ in terms of two real angle parameters $\rho_{\mu\alpha}$ and $\omega_{\mu\alpha}$. Concomitantly, Fukutome transformed spin-orbitals become one-qubit Bloch sphere states in terms of $\rho_{\mu\alpha}$ and $\omega_{\mu\alpha}$. This decomposition leads to a natural QC encoding of END/QC for the chosen systems wherein each individual electron can be assigned to a single qubit. We adopt this natural encoding in this investigation and will present alternative formulations in terms of standard QC encodings [16] in a sequel soon.

Within the described framework, we derive all the END/QC expressions for one-electron heteronuclear diatomic molecules. We present various formulas to evaluate the END/QC spin-orbitals, wavefunction, total energy, matrix $\mathbf{M}$ and vector $\mathbf{V}$ in both analytical and QC programming forms. In addition, we present the encoded Hamiltonian and full END/QC equations of motion. To evaluate $\mathbf{M}$ and $\mathbf{V}$, we design and code five QC circuit (cf. Figs. 2-6) on the QC software development kit QISKIT[43]; these circuits are the END/QC versions of the VQS circuit[17]. For illustration's sake, we analyze the step-by-step functioning of the END/QC circuit depicted in Fig. 3. With these circuits, we evaluate the elements/components of $\mathbf{M}$ and $\mathbf{V}$, and the whole $\mathbf{M}$ and $\mathbf{V}$ constructs, and calculate various metrics to gauge the accuracy (as MAE or mean RMSE) and precision (as SDs) of the QC results with respect to their analytical counterparts. Remarkably, in log$_2$-log$_2$ plots, we find that the error and deviation of all the QC results exhibit a perfect decreasing linear correlation with the number of shots $n_s$ and with a same slope equal to -1/2 (cf. Figs 7-13). In other words, we find that the error and deviation of these QC calculations scale asymptotically with order $O(n_s^{-1/2})$.

We illustrate a full END/QC/VQS simulation with the pure electronic dynamics of a $H_2^+$ molecule, a homonuclear diatomic subcase. We first demonstrate that this two-parameter END/QC dynamics has the analytic solution $\rho_{\mu\alpha}(t) = \rho_{\mu\alpha}^0 =$ constant and $\omega_{\mu\alpha}(t) = -\Delta_{\mu\alpha} t + \omega_{\mu\alpha}^0$, where the angular frequency $\Delta_{\mu\alpha}$ is the HOMO-LUMO energy gap. We exemplify the three main tasks of the END/QC/VQS algorithm with this $H_2^+$ case. Finally, we plot the time evolutions of the one-electron density and of the electron Mulliken populations corresponding to the END/QC/VQS dynamics of $H_2^+$. Both properties evolutions reveal a periodic intramolecular electron transfer between the two nuclei. The analytical solution of the END/QC equations indicates that the "trigger" of this dynamics is the breaking of the $D_{\infty h}$ spatial symmetry in the initial END/QC wavefunction; such breaking is generated by the superposition of the HOMO and LUMO that correspond to the *gerade* and *ungerade* 1-D irreducible representation of $D_{\infty h}$, respectively [19,40,41]. The plot of the electron Mulliken population vs. time from the numerically integrated END/QC/VQS equations agrees with its counterpart from analytical formulas. Those electron Mulliken populations oscillate in time with an angular frequency equal to the HOMO-LUMO energy gap $\Delta_{\mu\alpha}$.

This manuscript presents a successful *proof of concept* of END/QC/VQS, a solid foundation from which we can continue developing this method to its full maturity. Thus, in a series of forthcoming publications, we will first apply the current END/QC/VQS to larger model systems in order to further appraise this novel approach. Then, and more importantly, we will generalize the current END/QC/VQS for full electronic and nuclear dynamics, and for any type of molecules described with *ab initio* Hamiltonians and large basis sets. Finally, we will extend our QC formulation to the currently classical tasks I and III of END/QC/VQS.

**Author Contributions:** Conceptualization, IDF and JAM; methodology, IDF and JAM; software, JCD; validation, JCD and JAM; formal analysis, JCD and JAM; investigation, JCD and JAM.; data curation, JCD; writing—original draft preparation, JAM; supervision, IDF and JAM; project administration, JAM; funding acquisition, JAM.

**Acknowledgments:** All present calculations have been performed at the TTU High Performance Computer Center. This research was funded by the National Science Foundation Graduate Research Fellowship Program (DGE 2140745, to JCD), by a Texas Tech University (TTU) Graduate School Fellowship (to JCD), and by the National Institutes of Health (NIH) grant 1R15GM128149-01 (to JAM).

**Dedication:** We warmly dedicate this article to the memory of our collaborator and friend Professor Ismael de Farias, who encouraged us to enter the enthralling field of QC and taught us many useful things in that area.

**Table 1**: Values of the circuit parameters $\alpha$ and $a$, and of the END/QC variational parameters $\rho = \rho_{\mu\alpha}$ and $\omega = \omega_{\mu\alpha}$ for the QC evaluation of the elements $\tilde{M}_{jk}$ and $\tilde{V}_{j,k}$, and of the matrix **M** and vector **V** with the circuits in Figs. 2-6.

| Circuit-Evaluated Elements, Matrix and Vector | END/QC Parameters | | Value of the elements | Circuit Parameters | |
|---|---|---|---|---|---|
| | $\rho$ (°) | $\omega$ (°) | | $\alpha$ (°) | $a$ (a.u.) |
| $\tilde{M}_{23} = -a\sin(2\rho)$ | 240 | 180 | $\tilde{M}_{23} = -\sqrt{3}/2$ | 90 | 1.0 |
| $\tilde{V}_{2,X} = a\cos(2\rho)\cos(\omega)$ | 240 | 45 | $\tilde{V}_{2,X} = -\sqrt{2}/4$ | 90 | $2h_{\mu\alpha} = 2.0$ |
| $\tilde{V}_{2,Z} = -a\sin(2\rho)$ | 240 | 180 | $\tilde{V}_{2,Z} = -\sqrt{3}/2$ | 90 | $(h_{\alpha\alpha} - h_{\mu\mu}) = -0.7773$ |
| $V_{3,X} = -a\sin(2\rho)\sin(\omega)$ | 240 | 45 | $V_{3,X} = -\sqrt{6}/4$ | 90 | $h_{\mu\alpha} = 1.0$ |
| **M** Matrix | 240 | 180 | From previous circuits | Ibid. | Ibid. |
| **V** vector | 240 | 180 | Ibid. | Ibid. | Ibid. |

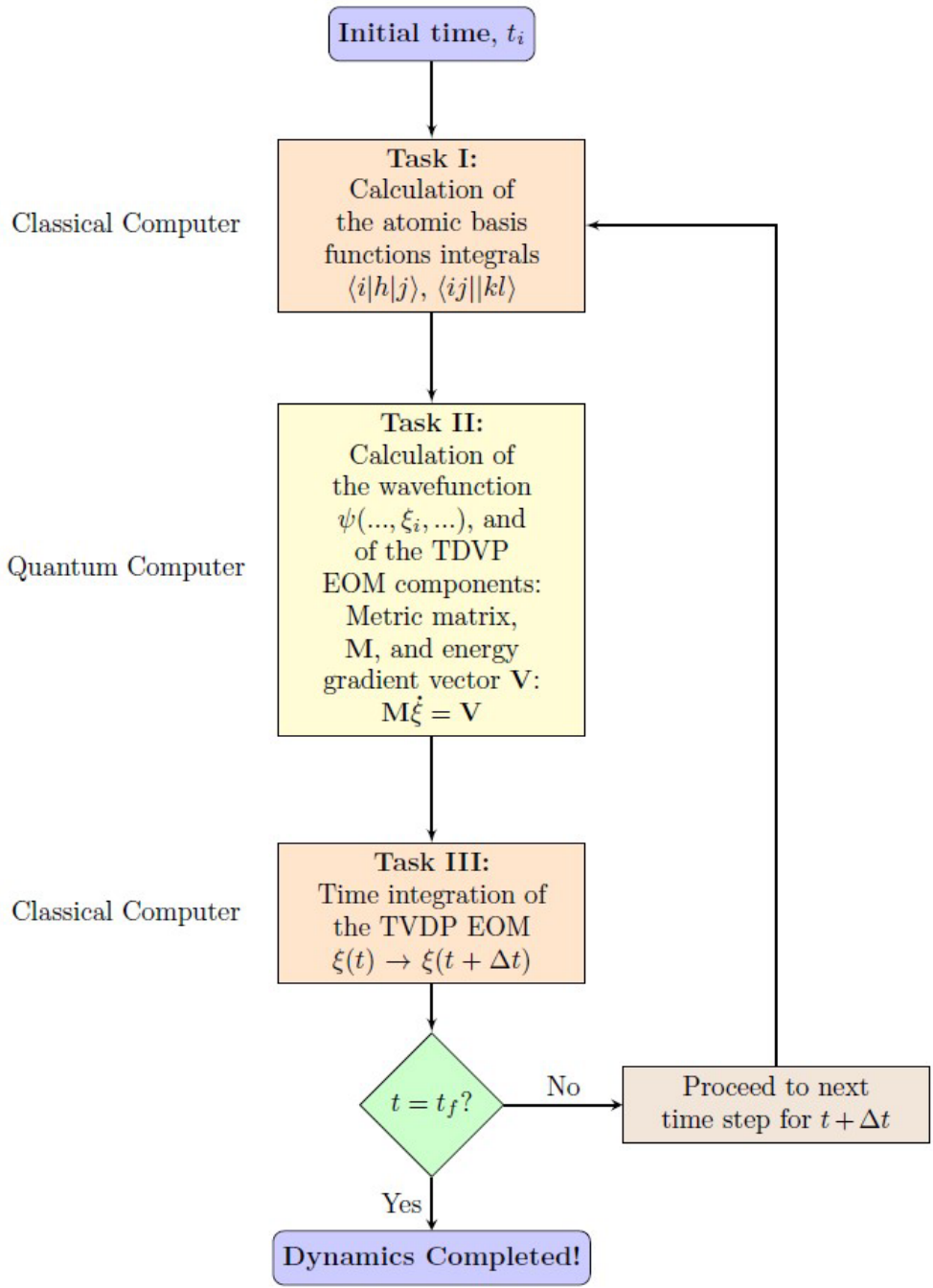

**Figure 1:** Flowchart of the main computational tasks in an END/QC/VQS dynamics executed on classical and quantum computers.

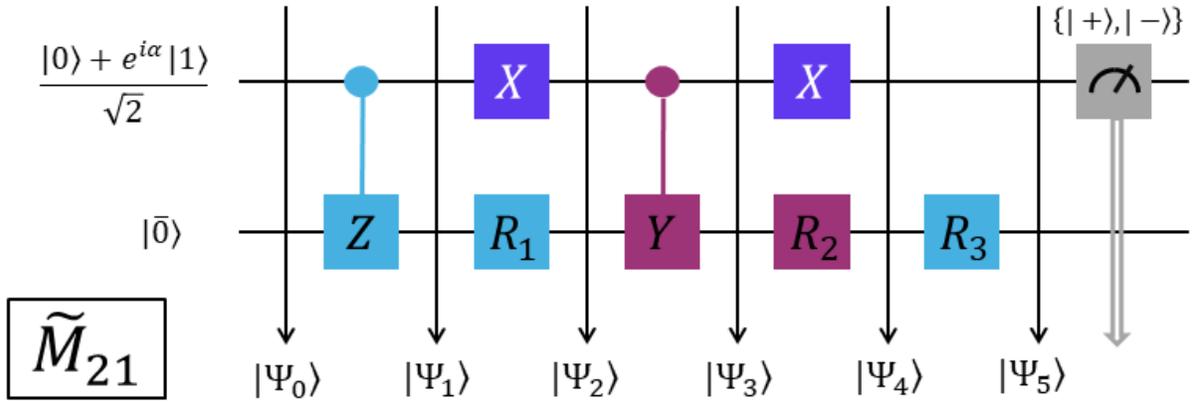

**Figure 2:** QC circuit to evaluate the element $\widetilde{M}_{21}$, Eq. (47).

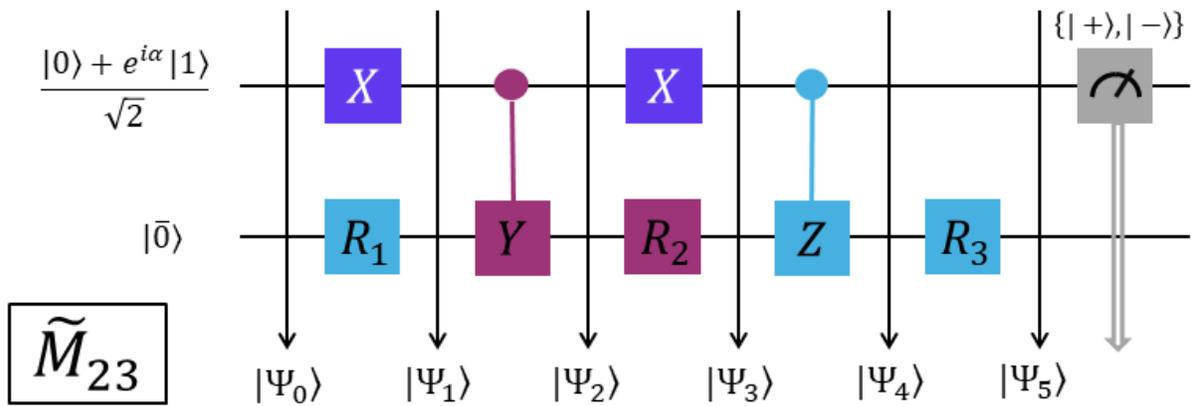

**Figure 3:** QC circuit to evaluate the element $\widetilde{M}_{23}$, Eq. (47).

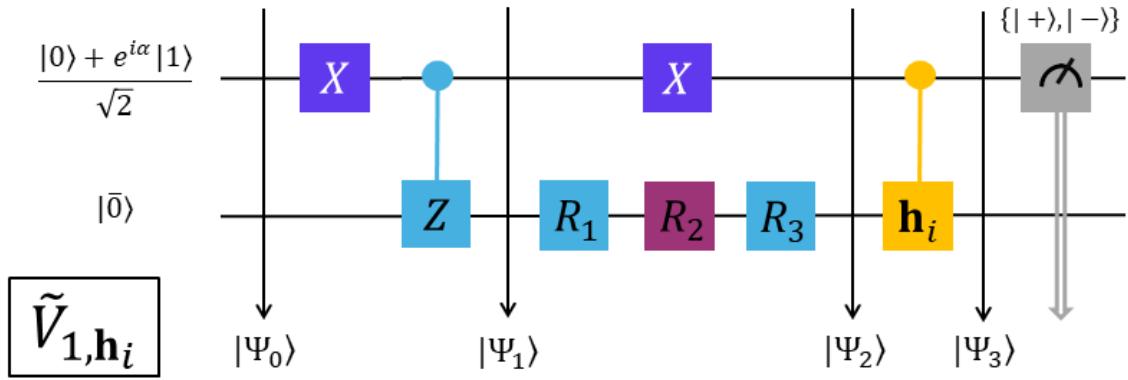

**Figure 4:** QC circuit to evaluate the element $\tilde{V}_{1,\mathbf{h}_i}$, $\mathbf{h}_i = k = \mathbf{I}, \mathbf{X}, \mathbf{Y}, \mathbf{Z}$, Eq. (48).

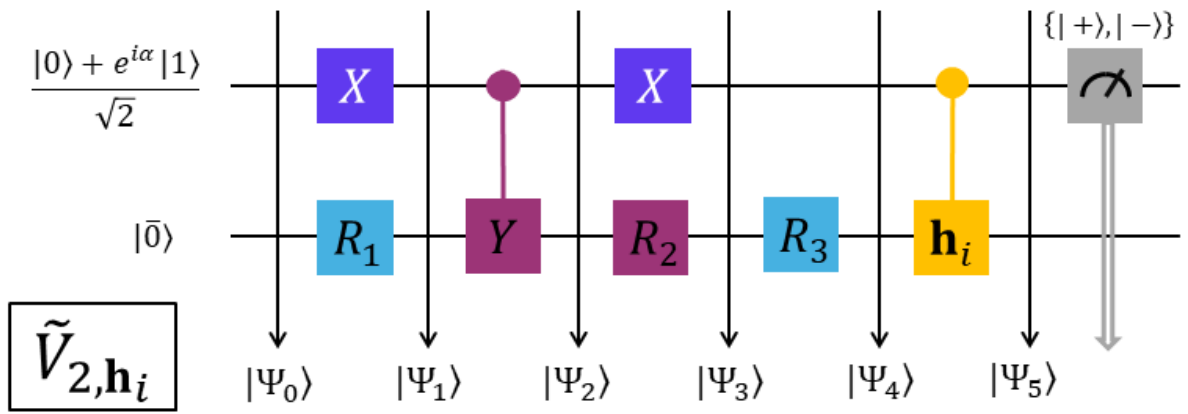

**Figure 5:** QC circuit to evaluate the element $\tilde{V}_{2,\mathbf{h}_i}$, $\mathbf{h}_i = k = \mathbf{I}, \mathbf{X}, \mathbf{Y}, \mathbf{Z}$, Eq. (48).

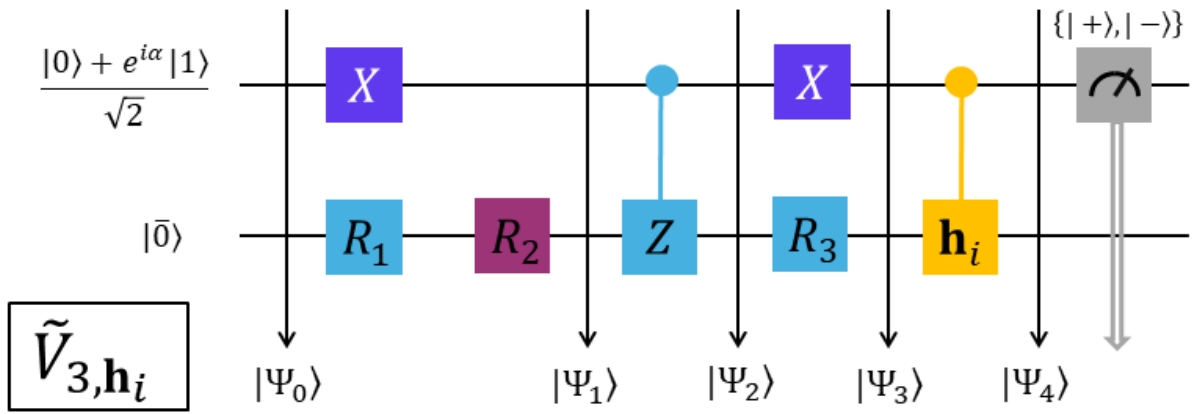

**Figure 6:** QC circuit to evaluate the element $\tilde{V}_{3,\mathbf{h}_i}$, $\mathbf{h}_i = k = \mathbf{I}, \mathbf{X}, \mathbf{Y}, \mathbf{Z}$, Eq. (48).

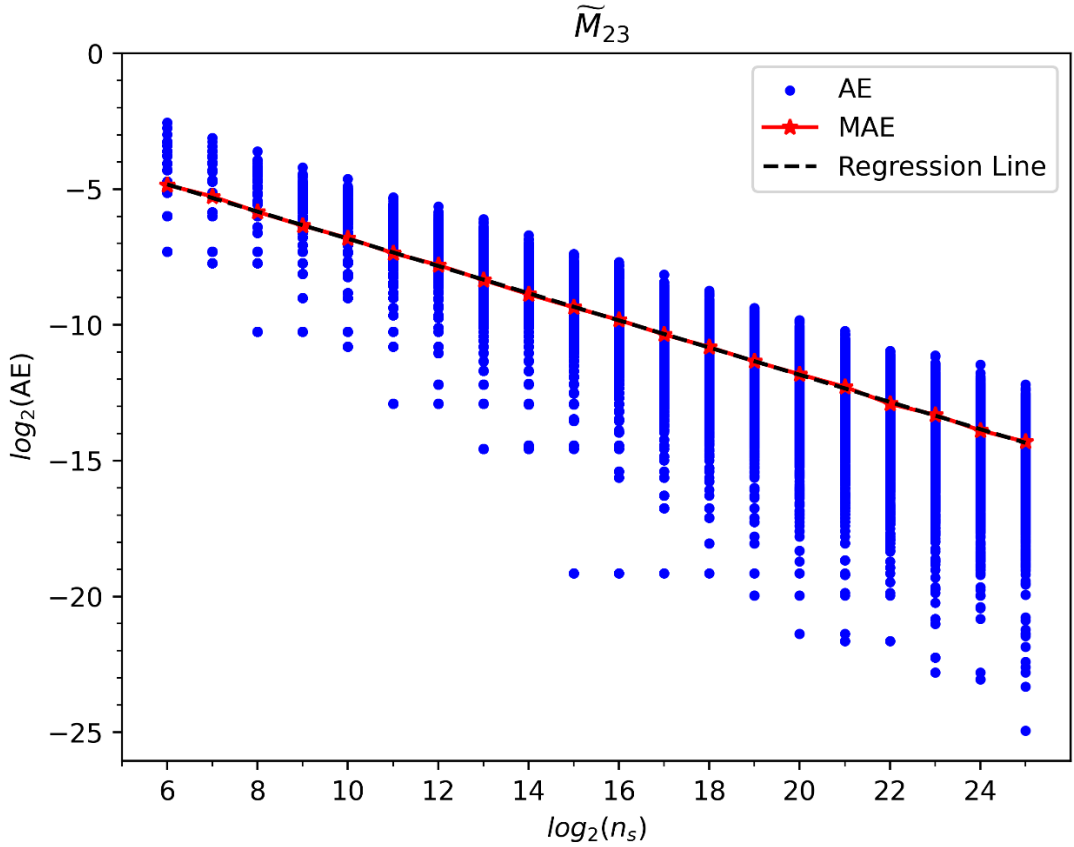

**Figure 7:** $log_2$-$log_2$ plot of the absolute error (AE) in QC calculations of the element $\widetilde{M}_{23}$ vs. the number of shots $n_s$. Values corresponding to END/QC variational parameters $\rho = 240°$ and $\omega = 180°$, and circuit parameters $\alpha = 90°$ and $a = 1$. AEs from 1,000 repetitions per each $n_s$ appear as scattered blue dots, and their mean absolute error (MAE) per each $n_s$ appear as red stars. For the latter, a regression line with slope $\tilde{\alpha} = -0.50077038$, intercept $\tilde{\beta} = -1.8216273$, and coefficient of determination $R^2 = 0.9999$ appears as a black dashed line.

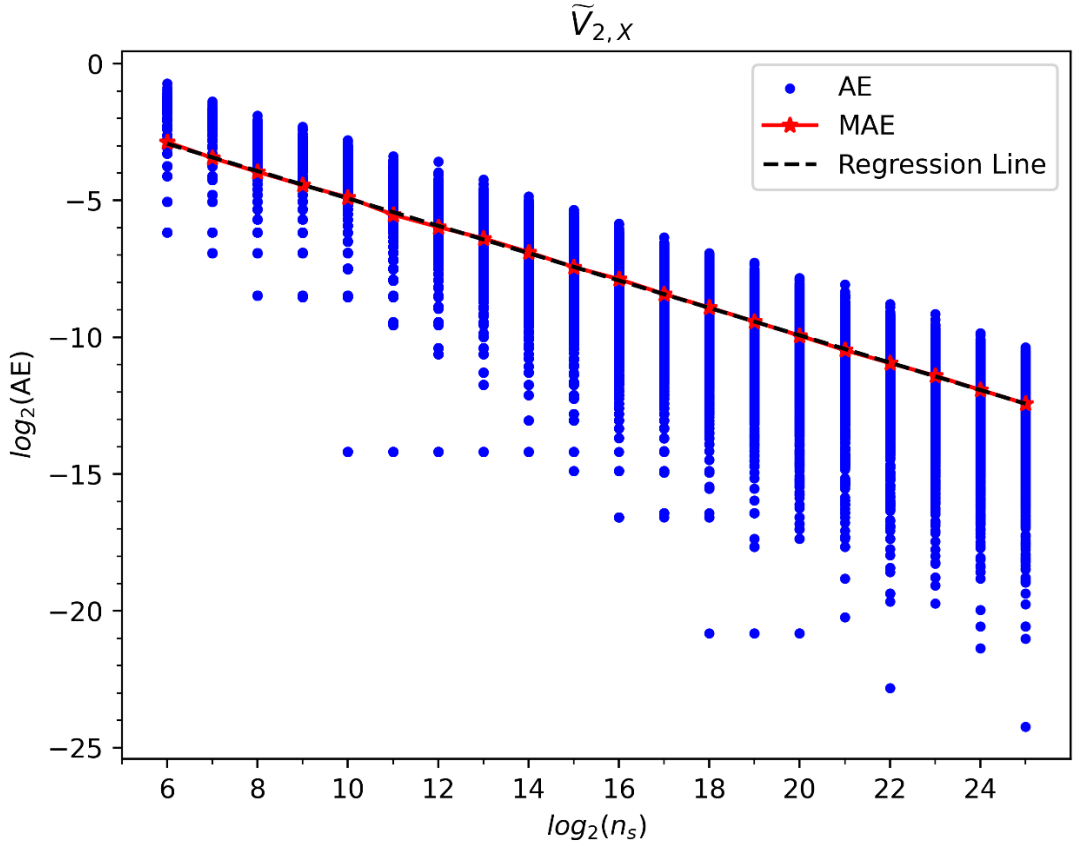

**Figure 8:** log$_2$-log$_2$ plot of the absolute error (AE) in QC calculations of the element $\widetilde{V}_{2,X}$ vs. the number of shots $n_s$. Values corresponding to END/QC variational parameters $\rho = 240°$ and $\omega = 45°$, and circuit parameters $\alpha = 90°$ and $a = 2\tilde{h}_X = 2h_{\mu\alpha} = 2.0$ a.u. AEs from 1,000 repetitions per each $n_s$ appear as scattered blue dots, and their mean absolute error (MAE) per each $n_s$ appear as red stars. For the latter, a regression line with slope $\tilde{\alpha} = -0.50006339$, intercept $\tilde{\beta} = 0.07408203$, and coefficient of determination $R^2 = 0.9998$ appears as a black dashed line.

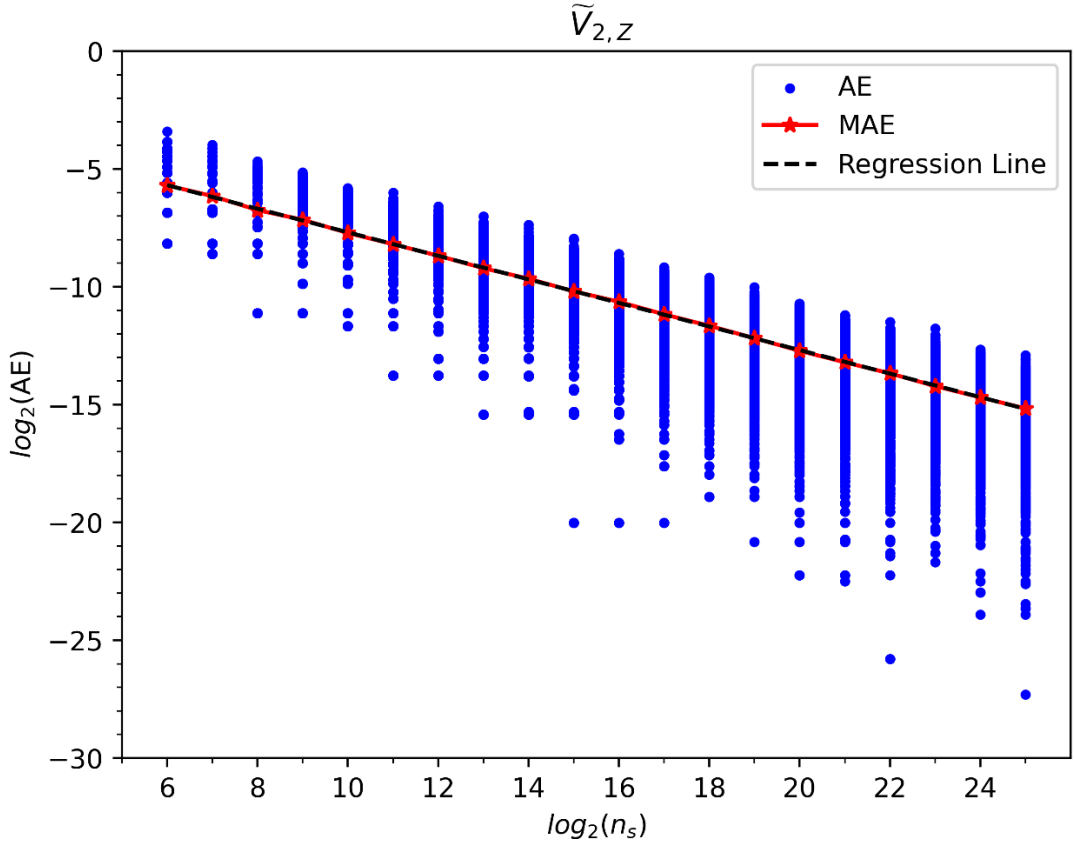

**Figure 9:** log$_2$-log$_2$ plot of the absolute error (AE) in QC calculations of the element $\tilde{V}_{2,z}$ vs. the number of shots $n_s$. Values corresponding to END/QC variational parameters $\rho = 240°$ and $\omega = 180°$, and circuit parameters $\alpha = 90°$ and $a = 2\tilde{h}_z = (h_{\alpha\alpha} - h_{\mu\mu}) = -0.7773$ a.u. AEs from 1,000 repetitions per each $n_s$ appear as scattered blue dots, and their mean absolute error (MAE) per each $n_s$ appear as red stars. For the latter, a regression line with slope $\tilde{\alpha} = -0.49975525$, intercept $\tilde{\beta} = -2.68894878$, and coefficient of determination $R^2 = 0.9999$ appears as a black dashed line.

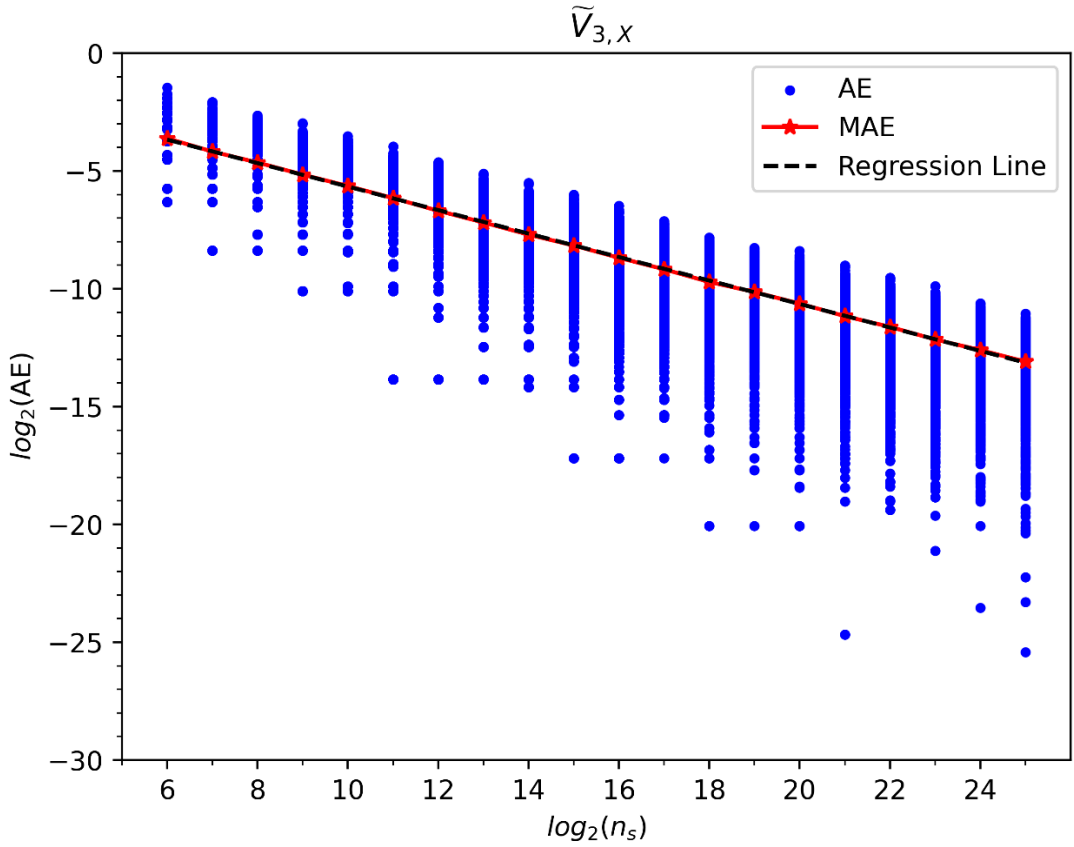

**Figure 10:** log₂-log₂ plot of the absolute error (AE) in QC calculations of the element $\widetilde{V}_{3,X}$ vs. the number of shots $n_s$. Values corresponding to END/QC variational parameters $\rho = 240°$ and $\omega = 45°$, and circuit parameters $\alpha = 90°$ and $a = \tilde{h}_X = h_{\mu\alpha} = 1$ a.u. AEs from 1,000 repetitions per each $n_s$ appear as scattered blue dots, and their mean absolute error (MAE) per each $n_s$ appear as red stars. For the latter, a regression line with slope $\tilde{\alpha} = -0.49819948$, intercept $\tilde{\beta} = -0.68106768$ and coefficient of determination $R^2 = 0.9999$ appears as a black dashed line.

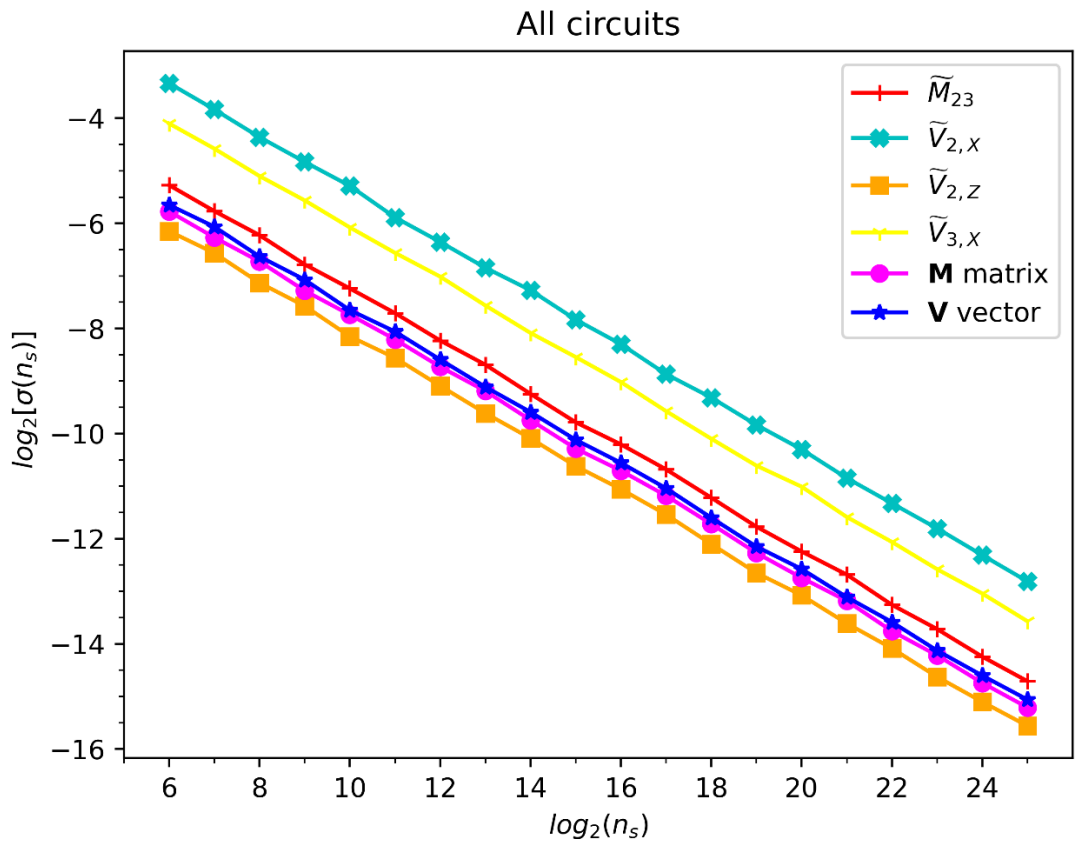

**Figure 11:** $\log_2[\sigma(n_s)]$ as a function of $\log_2(n_s)$ for the results of the individual circuits and of the **M** matrix and **V** vector. Legend on the upper right specifies the color and marker for each of the curves.

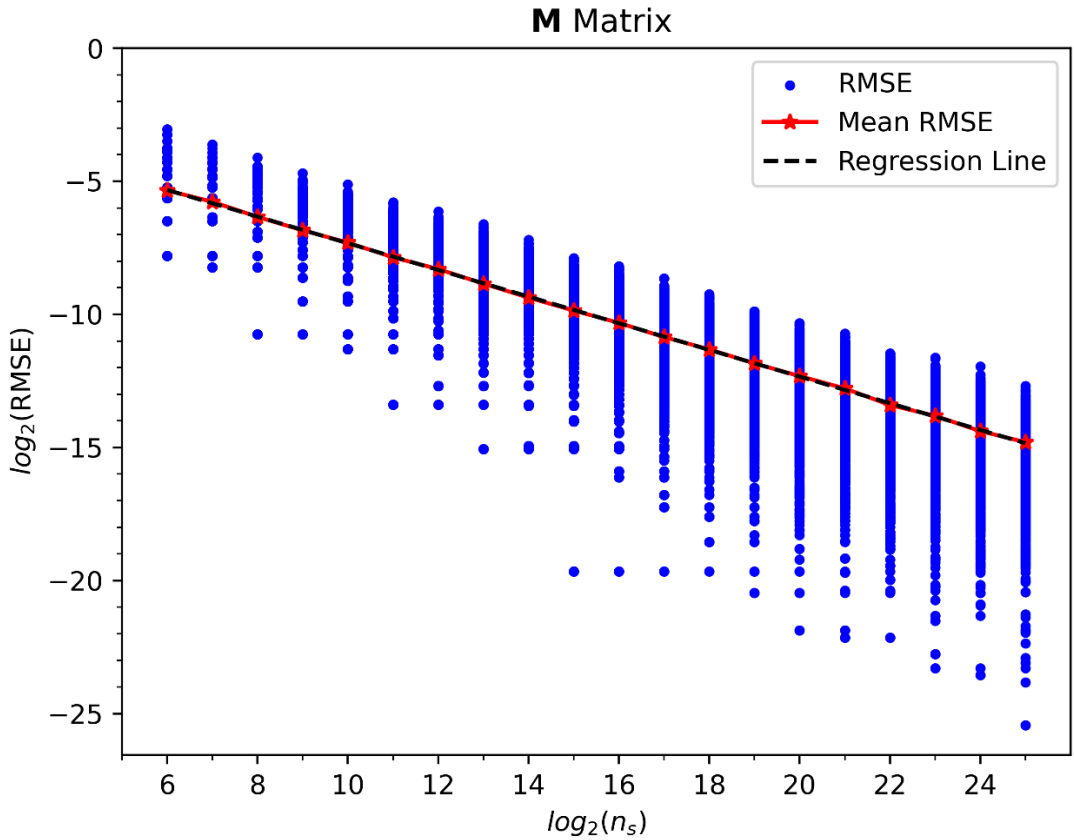

**Figure 12:** $\log_2$-$\log_2$ plot of the root mean square error (RMSE) in QC calculations of the matrix $\boldsymbol{M}$ vs. the number of shots $n_s$. Values corresponding to END/QC variational parameters $\rho = 240°$ and $\omega = 180°$, and circuit parameters $\alpha = 90°$ and $a = 1$. RMSEs from 1,000 repetitions per each $n_s$ appear as scattered blue dots, and their mean RMSEs per each $n_s$ appear as red stars. For the latter, a line of regression with slope $\tilde{\alpha} = -0.50077038$, intercept $\tilde{\beta} = -2.3216273$, and coefficient of determination $R^2 = 0.9999$ appears as a black dashed line.

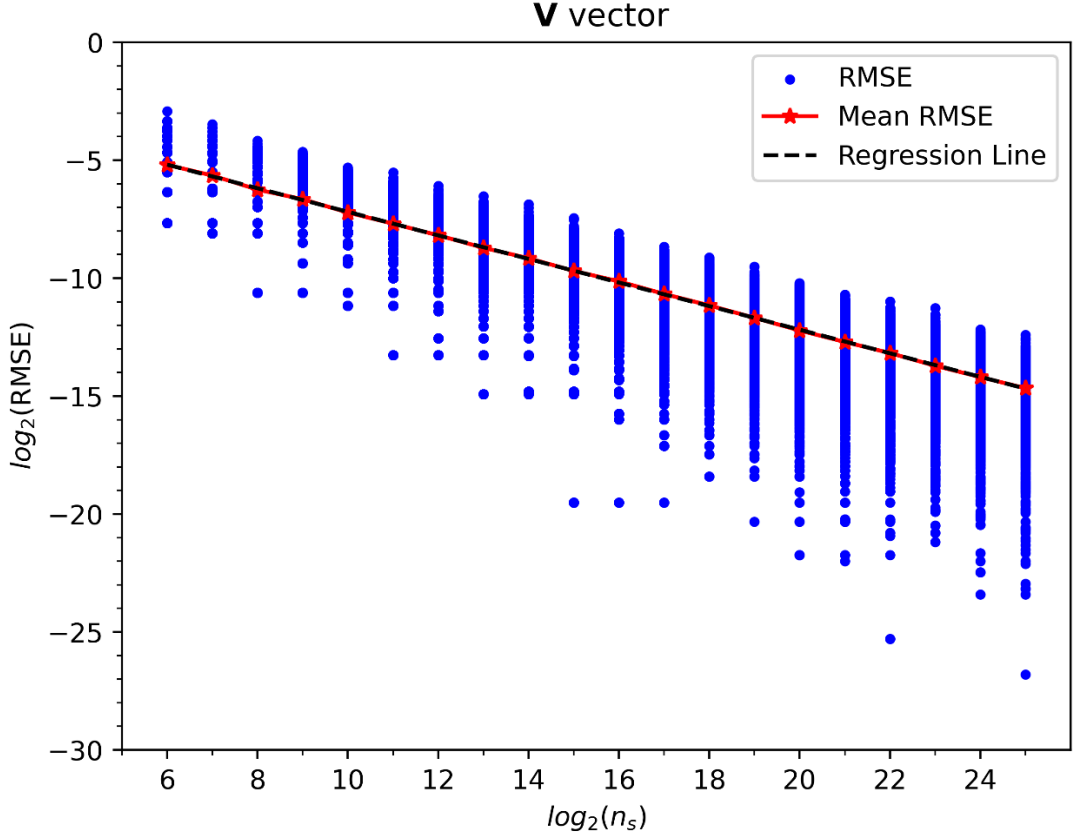

**Figure 13:** log$_2$-log$_2$ plot of the root mean square error (RMSE) in QC calculations of the vector **V** vs. the number of shots $n_s$. Values corresponding to END/QC variational parameters $\rho = 240°$ and $\omega = 180°$, and circuit parameters $\alpha = 90°$ and $a = 2\tilde{h}_z = (h_{\alpha\alpha} - h_{\mu\mu}) = -0.7773$ a.u. RMSEs from 1,000 repetitions per each $n_s$ appear as scattered blue dots, and their mean RMSEs per each $n_s$ appear as red stars. For the latter, a regression line with slope $\tilde{\alpha} = -0.49975525$, intercept $\tilde{\beta} = -2.18894878$, and coefficient of determination $R^2 = 0.9999$ appears as a black dashed line.

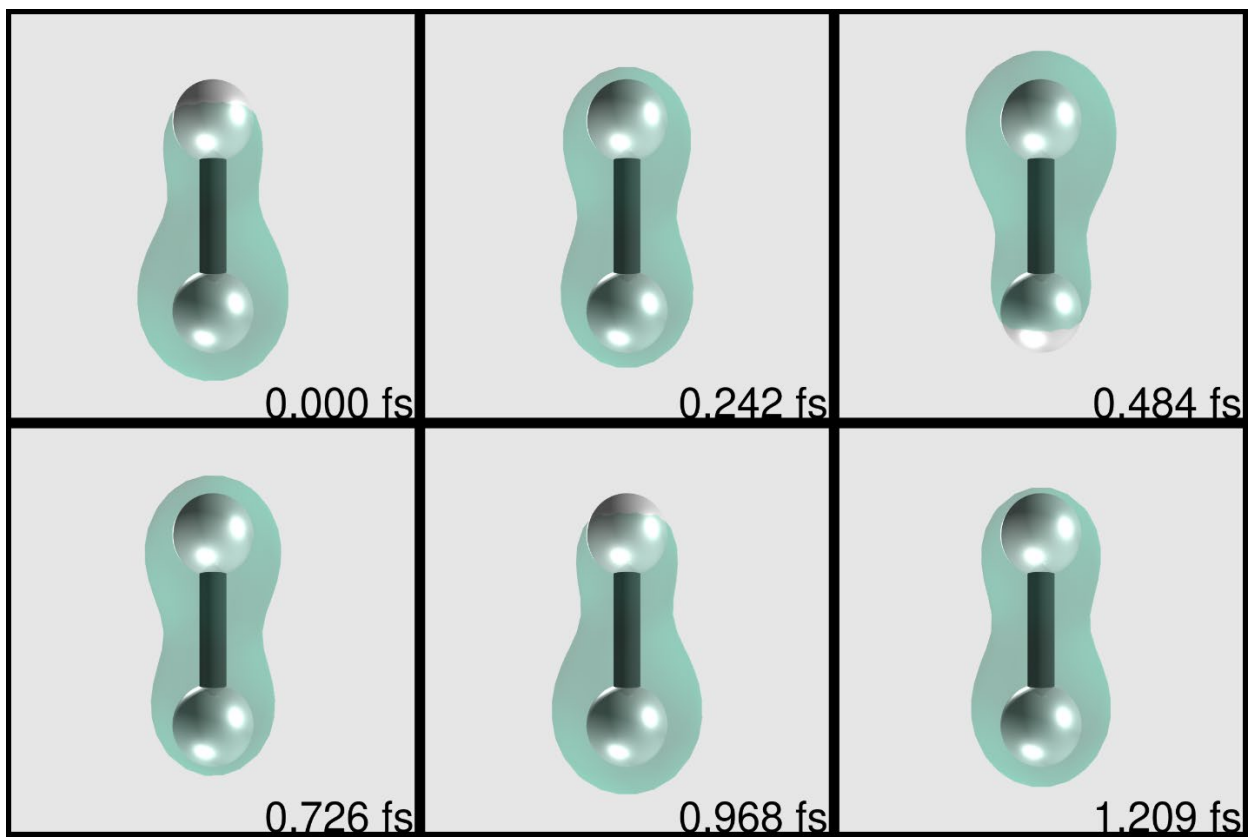

**Figure 14:** Six sequential snapshots of an END/QC/VQS/STO-3G computer animation of the pure electronic dynamics of $H_2^+$ at bond distance $R = 1.4$ a.u. and from the initial conditions $\rho_{\mu\alpha}^0 = 5^0$ and $\omega_{\mu\alpha}^0 = 0^0$. Simulation times in femtoseconds. The white spheres represent the fixed H nuclei and the green cloud depicts an electron density $\rho(\mathbf{r},t)$ iso-value = 0.1 a.u.

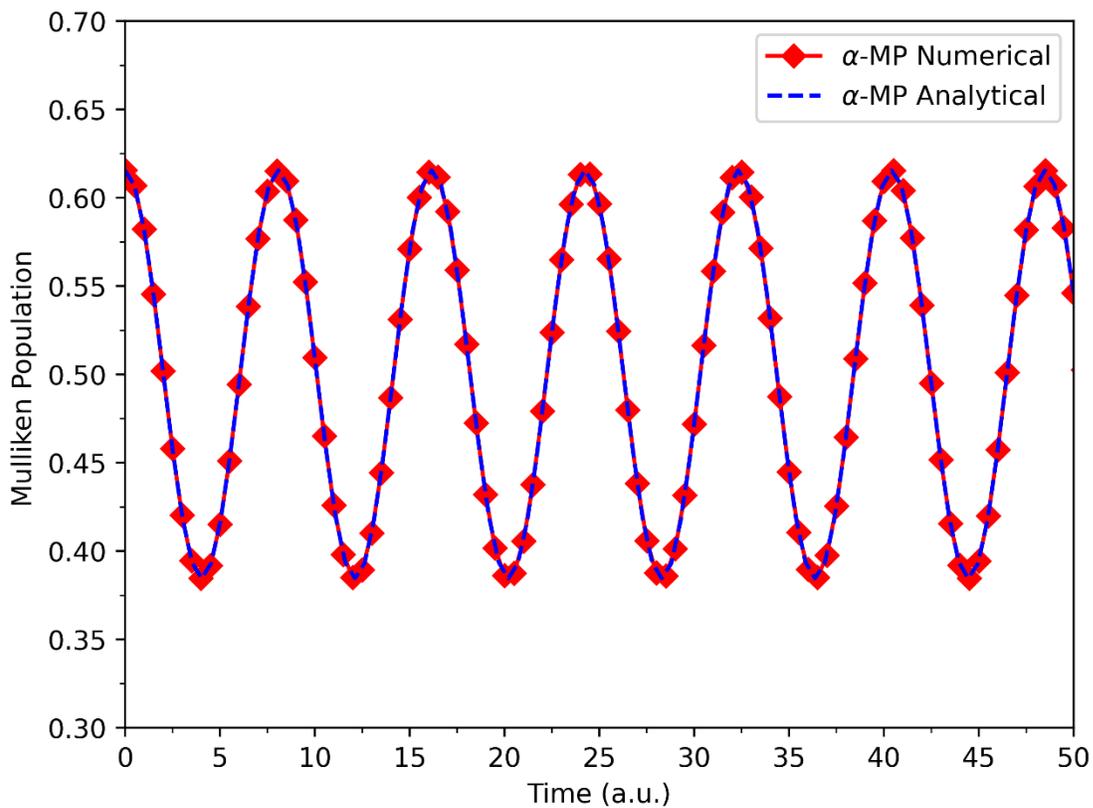

**Figure 15:** Electron Mulliken population (MP) on a H nucleus vs. time from an END/QC/VQS/STO-3G simulation of the pure electronic dynamics of $H_2^+$ at bond distance $R$ = 1.4 a.u. and from initial conditions $\rho_{\mu\alpha}^0 = 5^0$ and $\omega_{\mu\alpha}^0 = 0^0$.